\documentclass[%
 prx,
 jmp,%
 amsmath,amssymb,
 floatfix,
 reprint,%
 showpacs,%
]{revtex4-2}

\usepackage{comment}
\newcommand{\cmmnt}[1]{\ignorespaces}
\usepackage{amssymb}
\usepackage{amsmath}
\usepackage{mathtools}
\usepackage{relsize}
\usepackage{array}
\usepackage{bm}
\usepackage{graphicx}
\usepackage{bigints}
\usepackage{ifthen}
\usepackage{multirow}
\usepackage{threeparttable}
\usepackage{rotating}
\usepackage{hyperref}
\usepackage{xcolor}
\usepackage{calrsfs}
\usepackage{empheq}
\usepackage[normalem]{ulem}
\usepackage{media9}
\usepackage[margin=0pt,font=small,labelfont=normalfont,labelsep=period,format=plain,justification=centerlast]{caption}

\DeclareGraphicsExtensions{.pdf,.png,.eps}
\graphicspath{{figs/}{./}}
\usepackage{cleveref}
\crefname{equation}{}{}
\Crefname{equation}{}{}
\crefname{figure}{Fig.~\!\!\!}{Figs.~\!\!\!}
\Crefname{figure}{Fig.~\!\!\!}{Figs.~\!\!\!}

\usepackage{tikz}
\usetikzlibrary{intersections,arrows,decorations,decorations.pathmorphing,calc}
\usepackage{pgfplots}
\pgfplotsset{compat=newest}
\pgfplotsset{plot coordinates/math parser=false}

\makeatletter
\let\jnl@style=\rmfamily
\def\ref@jnl#1{{\jnl@style#1}}
\@ifundefined{aap}{\newcommand\aap{\ref@jnl{A\&A}}}{}%
\@ifundefined{apj}{\newcommand\apj{\ref@jnl{ApJ}}}{}%
\@ifundefined{apjl}{\newcommand\apjl{\ref@jnl{Astrophys.~J.~Lett.}}}{}%
\@ifundefined{ssr}{\newcommand\ssr{\ref@jnl{Space~Sci.~Rev.}}}{}%
\@ifundefined{jgr}{\newcommand\jgr{\ref@jnl{J.~Geophys.~Res.}}}{}%
\@ifundefined{jatp}{\newcommand\jatp{\ref@jnl{J.~Atmos.~Terr.~Phys.}}}{}%
\@ifundefined{solphys}{\newcommand\solphys{\ref@jnl{Sol.~Phys.}}}{}%
\@ifundefined{pp}{\newcommand\pp{\ref@jnl{Phys.~Plasmas}}}{}%
\@ifundefined{mnras}{\newcommand\mnras{\ref@jnl{MNRAS}}}{}%
\@ifundefined{apjs}{\newcommand\apjs{\ref@jnl{ApJS}}}{}%
\@ifundefined{azh}{\newcommand\azh{\ref@jnl{AZh}}}{}%
\catcode`\@=11

\newcommand{\abs}[1]{\left| #1 \right|}

\newcommand{\lims}[1]{\left[ #1 \right]}
\newcommand{\paren}[1]{\left( #1 \right)}
\newcommand{\pfrac}[2]{\left( \frac{#1}{#2} \right)}

\def\beq#1\eeq{\begin{equation}#1\end{equation}}
\def\bes#1\ees{\begin{subequations}#1\end{subequations}}
\def\bea#1\eea{\begin{align}#1\end{align}}
\def\nn{\nonumber\\}

\newcommand{\D}[2]{\frac{\partial #1}{\partial #2}}

\newcommand{\mvec}[1]{\bm{#1}}

\def\sscomma{\, , \,}
\def\o{\omega}

\usepackage{scalerel}

\allowdisplaybreaks

\usepackage{soul}
\usepackage{media9}
\usepackage{hyperref}
\graphicspath{{./}{figs/}{/scratch/vit/Data/BW/BW-img/}}
\DeclareGraphicsExtensions{.pdf,.png,.eps}
\usetikzlibrary{positioning}
\newlength{\fw}
\newlength{\ws}
\def\fbt{\fbox{\hspace*{0.975\ws}\vbox to 0.975\ws{}}}
\usepgfmodule{oo}
\pgfooclass{tmplt} {
  \method tmplt() { }
  \method apply(#1,#2,#3,#4,#5,#6) {
    \fill [black!0!white] (-2.55, 2.55) rectangle (2.55,-7.55);
    \node[inner sep=0pt,opacity=1.0] (v00) at (0,0)%
      {\includemedia[playbutton=none,width=\ws,flashvars={source=#1&autoPlay=true&loop=true}] {
        \includegraphics[angle=0,width=\ws,clip,trim=150 150 150
          150,decodearray={0 1 0 1 0 1 0 1}]{#2}}{VPlayer9.swf}}; 
    \node[black] at (0,0) {\href{#3}{\fbt}};
    \node[inner sep=0pt,opacity=1.0] (v01) at (0,-5)%
      {\includemedia[playbutton=none,width=\ws,flashvars={source=#4&autoPlay=true&loop=true}] {
        \includegraphics[angle=0,width=\ws,clip,trim=150 150 150
          150,decodearray={0 1 0 1 0 1 0 1}]{#5}}{VPlayer9.swf}}; 
    \node[black] at (0,-5) {\href{#6}{\fbt}};
    \node[inner sep=0pt,opacity=1.0] at (0,2.3) {\bf Complete wave trajectory};
    \node[inner sep=0pt,opacity=1.0] at (0,-2.7) {\bf Emergent loop pattern};
  }
}
\usepackage{siunitx}
\newcommand{\der}[2]{\frac{d {#1}}{d {#2}}}
\newcommand{\overbar}[1]{\mkern 1.5mu\overline{\mkern-1.5mu#1\mkern-1.5mu}\mkern 1.5mu}
\newcommand{\dl}{\overbar{\delta}}

\usepackage[inline]{trackchanges}
\addeditor{LF}
\addeditor{VG}
\soulregister\cite7

\begin{document}

\title{The Nature of the Action Potential}

\author{Vitaly L. Galinsky}
\email{vit@ucsd.edu}
\affiliation{Center for Scientific Computation in Imaging,
University of California at San Diego, La Jolla, CA 92037-0854, USA}
\author{Lawrence R. Frank}
\email{lfrank@ucsd.edu}
\affiliation{Center for Scientific Computation in Imaging,
University of California at San Diego, La Jolla, CA 92037-0854, USA}
\affiliation{
Center for Functional MRI,
University of California at San Diego, La Jolla, CA 92037-0677, USA}

\date{\today}

\begin{abstract}

We demonstrate that our recently developed theory of electric field wave propagation in anisotropic and inhomogeneous brain tissues, which has been shown to explain a broad range of observed coherent synchronous brain electrical processes, also explains the spiking behavior of single neurons, thus bridging the gap between the fundamental element of brain electrical activity (the neuron) and large-scale coherent synchronous electrical activity. 

Our analysis indicates that the membrane interface of the axonal cellular system can be mathematically described by a nonlinear system with several small parameters. This allows for the rigorous derivation of an accurate yet simpler nonlinear model following the formal small parameter expansion. The resulting action potential model exhibits a smooth, continuous transition from the linear wave oscillatory regime to the nonlinear spiking regime, as well as a critical transition to a non-oscillatory regime. These transitions occur with changes in the criticality parameter and include several different bifurcation types, representative of the various experimentally detected neuron types. 

This new theory overcomes the limitations of the Hodgkin-Huxley model, such as the inability to explain extracellular spiking, efficient brain synchronization, saltatory conduction along myelinated axons, and a variety of other observed coherent macroscopic brain electrical phenomena. We also show that the standard cable axon theory can be recovered by our approach, using the very crude assumptions of piece-wise homogeneity and isotropy. However, the diffusion process described by the cable equation is not capable of supporting action potential propagation across a wide range of experimentally reported axon parameters.
\end{abstract}

\maketitle

\section{Introduction}
The Hodgkin and Huxley model for axonal electric signaling
\cite{pmid12991237} is a cornerstone of modern neuroscience and
serves as the basis for the development of a wide range of complex
models of brain electrical communication.  This model, and a host of
variations (e.g.,
\cite{pmid19431309,Nagumo1962,pmid7260316,pmid18244602}), (hereafter
collectively referred to as HH) is based on the postulate that axons
possess multiple voltage gated channels that open and close in
synchrony thereby producing a coherent persistent electrical wave or
'spike' traveling along the axon.  In order to parameterize
experimentally observed spikes in support of this view, HH
developed a model described by a reaction-diffusion process.
However, despite the general utility, and universal acceptance of
this model, several incontrovertible facts suggest its
incompatibility with observed brain electrical activity, such as its
inability to account for extracellular spiking, efficient brain
synchronization, and saltatory conduction along myelinated axons, to
name just a few.

In retrospect, this should not be surprising, as the HH model
was never derived from first physical principles, but was an
\textit{ad hoc} construction based on a very simple model motivated
more by its flexibility than its adherence to any first physical
principles of electrodynamics.  After all, a
general multiparametric reaction-diffusion equation constructed
with multiple time constants, thresholds, and power laws can
empirically fit a multitude of physical systems, including the
hypothesized neuron with multiple voltage gated channels, even if it is
not the correct physical model.  That trouble was brewing should
have been evident from the fact that this asynchronous, seemingly
incoherent spiking activity at scales of a single neuron appeared
inconsistent with observed oscillatory and wave-like patterns that
are coherent across a wide range of spatial and temporal scales
\cite{buzsaki2006rhythms}. Attempts to reconcile these seemingly
incompatible views led to the development of networks of
incoherently spiking neurons
\cite{Strassberg1993-nh,Meunier2002-rs,Yamazaki2022-cm}.  However,
because the original HH model is too complicated to describe even
relatively small networks, these networks models were modified to be
based on a very simplified but now ubiquitous model of a leaky
integrate-and-fire (LIF) neuron where a single threshold and time
constant replaces all the multiple gates, currents, channels and
thresholds
\cite{pmid19431309,Nagumo1962,pmid7260316,pmid18244602,Gerstner:2014:NDS:2635959,pmid33192427,pmid33288909}.
The unfortunate consequence is that, rather than reconciling the two
views, they now became incompatible, as LIF equations do not have a
mechanism for any type of non-linear resonance to generate the
sustained coherent traveling waves characteristic of neuronal
``spiking'' \cite{Galinsky:2022a}.

The source of these difficulties can be traced back to the lack of an
accurate physical model of electric field dynamics that includes wave
propagation and interaction in the anisotropic and inhomogeneous
neural tissues. In an effort to address this deficiency, we developed
such a theory which predicted the existence of previously undiscovered
weakly evanescent transverse cortical brain waves (WETCOW) generated
at surfaces (or interfaces) in neural tissues as a direct consequence
of their anisotropy and inhomogeneity.  This theory was shown to
describe a wide range of observed coherent macroscopic brain
electrical activity, including extracellular spiking, hypersynchronous
spiking and bursting, neuronal avalanches, and cortical wave loops
\cite{Galinsky:2020a,Galinsky:2020b,Galinsky:2021a,Galinsky:2021c}. However,
although the relationship to wave propagation in single neurons was
implicit in these papers, it was not demonstrated explicitly.  We do
so in this current paper by applying the WETCOW theory to an
analytical model of a single neuron with a lipid bilayer with an
anisotropic membrane conductivity.  The consequence is the generation
of waves of multiple frequencies and wave numbers propagating in the
lipid bilayer axonal membrane that create coherent nonlinear wave
states consistent with the spatial-temporal characteristics of
experimentally observed single neuron action potentials.

Further, having derived this directly from first principles
that incorporate tissue properties, we are able to directly predict
the well known observation that signals propagate faster along
myelinated axons, a result not attainable with the HH model.  From a
broader perspective, the demonstration that these coherent
persistent traveling non-linear waves are a consequence solely of
the electromagnetic properties of the neuronal tissues suggests that
it may be these waves that modify states of multiple cross-membrane
channels, causing them to open and close, rather the other way
around, which would be a fundamental shift in the understanding of
brain signaling.

\section{Theory}
The approach is similar to that developed in our general
theory
\cite{Galinsky:2020a,Galinsky:2020b,Galinsky:2021a,Galinsky:2021b,Galinsky:2021c,Galinsky:2022a,Galinsky:2023a,Galinsky:2023b}
and is as follows.  We begin with the general form of
electromagnetic activity (Maxwell's equations), from which we derive
the charge continuity equation in complex anisotropic and
inhomogeneous tissues.  This equation is then solved within a
cylindrical geometry representation of an idealized neuron with an
inhomogeneous and anisotropic membrane of finite thickness
surrounded on its inner and outer surfaces by a homogeneous
isotropically conducting fluids.  The key here is the inclusion of a
membrane conductivity tensor that provides a reasonable
approximation to the electrical properties of a lipid bilayer.  We
then solve the simple linear problem which demonstrates the
existence of surface waves even for this reduced solution.  We then
extend this to the more realistic non-linear problem and demonstrate
the existence of surface waves whose spatiotemporal characteristics
match those of observed data of neuronal spiking, though now derived
from first principles and thus directly related to neuronal geometry
and microstructure.

\subsection{The charge continuity equation}
In the most general form, a description of 
electromagnetic activity in axon can be formulated through Maxwell
equations in a medium, that are appropriate both for extracellular and
intracellular regions, 
\citep{Scott1975-ht,Bedard2004-lz}
\makeatletter%
\ifdim\textwidth>\columnwidth
  \setlength{\fw}{0.9\columnwidth}
  \begin{align}
  \nabla\cdot\mvec{D} &= \rho,\quad
  \nabla\times\mvec{H} = \mvec{J} + \D{\mvec{D}}{t}\quad \Rightarrow \quad
  \D{\rho}{t} &+ \mvec{\nabla}\cdot\mvec{J} = 0.\nonumber
  \end{align}
\else
  \setlength{\fw}{0.6\columnwidth}
  \begin{align}
  \nabla\cdot\mvec{D} &= \rho,\quad
  \nabla\times\mvec{H} = \mvec{J} + \D{\mvec{D}}{t}\quad \Rightarrow \quad
  \D{\rho}{t} + \mvec{\nabla}\cdot\mvec{J} = 0.\nonumber
  \end{align}
\fi
\makeatother

Using the electrostatic potential $\mvec{E}=-\nabla \phi$,
Ohm's law $\mvec{J}=\mvec{\sigma}\cdot\mvec{E}$ (where
$\mvec{\sigma}\equiv\{\sigma_{ij}\}$ is an anisotropic conductivity
tensor), a linear electrostatic property for brain tissue
$\mvec{D}=\varepsilon\mvec{E}$, assuming that the 
scalar permittivity $\varepsilon$ is a
``good'' function (i.e. it does not go to zero or infinity everywhere)
and making the change of variables $\partial x \to \varepsilon
\partial x^\prime$, the charge continuity equation for the
spatial-temporal evolution of the potential $\phi$ can be written in
terms of a permittivity scaled conductivity tensor
$\bm{\Sigma}=\{\sigma_{ij}/\varepsilon\}$ as
\begin{align}
\label{eq:phiSigma}
\D{}{t} \left(\nabla^2 \phi \right) &=
-\mvec{\nabla}\cdot\mvec{\Sigma}\cdot \nabla\phi + \mathcal{F}, 
\end{align}
where we have allowed for the influence of other sources by the
inclusion of a source (or forcing) term $\mathcal{F}$, that may have
both linear and nonlinear parts. This can be written in tensor
notation as
\begin{align}
\label{eq:phiTensor}
\partial_t \partial_i^2 \phi &+
\partial_i \left(\Sigma_{ij}\partial_j\phi\right)=0,
\end{align}
where repeating indices denotes summation. 

\subsection{The conductivity tensor of the lipid bilayer}
As shown in our earlier work \citep{Galinsky:2020a},
the existence of electric field surface waves is predicated
on the inhomogeneity and anisotropy of the neural tissues.
Remarkably, though, this does not require an exceedingly accurate
characterization of tissue microstructure. Rather, local average
tissue parameterizations are sufficient to make accurate predictions
of complex local and long-range non-linear wave propagation
properties.  This is an important point, as there is a huge body of
literature focused on suggesting the need for very accurate complex
tissue models to accurately predict observed coherent macroscopic
electromagnetic brain activity.  As demonstrated in our previous
publications, this is not the case
\cite{Galinsky:2020a,Galinsky:2020b,Galinsky:2021a,Galinsky:2021b,Galinsky:2021c,Galinsky:2022a,Galinsky:2023a,Galinsky:2023b}.

The same holds true for the single axon case considered here,
where a reasonable model for the membrane conductivity tensor
$\bm{\Sigma}^m$ can be constructed using a set of pretty general
assumptions and results in the generation of surface waves in
the lipid membrane. First, it is assumed that both the
along-axon (i.e., axial $z$) and across-axon (i.e., radial $r$) electric fields
will generate currents not only along the electric filed direction (i.e., along $z$ and $r$,
respectively) but will also generate currents that are perpendicular to the field (i.e., along $r$ and
$z$, respectively).  That is, for radial (along $r$) fields $\Sigma_{zr} \ne 0$ and
for axial (along $z$) fields $\Sigma_{rz} \ne 0$.  
Based only on these
symmetry considerations, the membrane conductivity tensor
$\bm{\Sigma}^m$ is assumed to have the following
non-diagonal, asymmetric form
\begin{align}
\label{eq:RZ}
\bm{\Sigma}^m &=
\begin{pmatrix}
\Sigma_{rr}(\phi) & \Sigma_{rz}(\phi)\\
\Sigma_{zr}(\phi) & \Sigma_{zz}(\phi)
\end{pmatrix},
\end{align}
where we indicate the fact that Ohm's law inside the membrane is
non-linear by adding a dependence of the conductivity tensor
components on scalar potential $\phi$.  The currents generated
parallel to the field are not expected to be equal (i.e. $\Sigma_{rr}
\ne \Sigma_{zz}$) nor would be equal the currents generated
perpendicular to the field ($\Sigma_{rz} \ne \Sigma_{zr}$).

\subsection{Axon model}
Both equations {\cref{eq:phiSigma}} and {\cref{eq:phiTensor}} are
appropriate for anisotropic and inhomogeneous media in general
geometry.  However, for the purposes of this paper it is sufficient to
consider an idealized model for an axon represented by a cylindrical
shell of diameter $d$ created by a membrane of thickness $\delta$
(that for myelinated axons includes the thickness of the myelin layers
as well) that separates two homogeneous isotropically conducting
fluids inside and outside of the shell with scaled conductivities
$\Sigma^i=\sigma_i/\varepsilon_i$ and
$\Sigma^e=\sigma_e/\varepsilon_e$. The conductivity inside the thin
membrane is highly anisotropic and is specified in tensor form
of the axially symmetric conductivity tensor $\bm{\Sigma}^m$
given by \cref{eq:RZ}.

Experimental measurements have shown that extracellular and
intracellular conductivities are similar to that of sea water ($\sim$4
S/m), or more exactly in the range from 0.28 S/m to 2.9 S/m for
extracellular $\sigma_e$ and intracellular $\sigma_i$
\citep{Scott1975-ht,Bedard2004-lz}, and permittivities $\varepsilon_e$
and $\varepsilon_i$ are around $7\times 10^{-10}$ F/m.  Thus both
$\Sigma^i$ and $\Sigma^e$ are very large, on the order of $10^{10}$
Hz.

The conductivity of the membrane is significantly smaller.  The values
vary and can be assumed to be in the range from as low as $10^{-13}$
S/m or as high as $10^{-5}$ \citep{Scott1975-ht}, with typical values
around $10^{-9}$ S/m \citep{Bedard2004-lz}. With comparable or
slightly smaller values for the membrane dielectric permittivity
$\varepsilon_m\sim 10^{-11}$ F/m it gives for the membrane scaled
conductivity $|\bm{\Sigma}|$ range estimate from $10^{-2}$ to $10^2$
Hz, hence the ratio of the conductivities of the membrane and 
the extracellular/intracellular media is as small as $10^{-8}$ to $10^{-12}$.

Because of this significant difference in scaled conductivities
between the membrane and the surrounding fluids, for the analysis of
electrodynamic processes near the membrane in the frequency range
characteristic of axonal signaling it can be reasonably assumed that
both extracellular and intracellular fluids act as very good (even
perfect) conductors that keep the potential drop across the membrane
at the resting potential value of $-V_0$ ($V_0\sim$65 mV).  This allows using
all variables normalized to the resting potential and scaled
conductivity tensor of the internal fluids.  Specifically, all the
variables in equations \cref{eq:phiSigma} and \cref{eq:phiTensor} are
normalized as $r\rightarrow r/d$, $\Sigma_{ij}\rightarrow
\Sigma_{ij}/\Sigma^i$, $t\rightarrow t\Sigma^i$, and
$\phi\rightarrow\phi/V_0$. We will also introduce normalized frequencies
($\omega\rightarrow \omega/\Sigma^i$) and radial ($\kappa\rightarrow
\kappa d$) and axial ($k\rightarrow k d$) wave numbers that will be
used later.  For the normalized membrane thickness ($\dl\rightarrow
\delta/d$) it will be assumed that $\dl < 1$. Often the difference
between $d$ and $\delta$ is significant so that $\dl \ll 1$.

\begin{figure}[!tbh] \centering
\includegraphics[width=\fw]{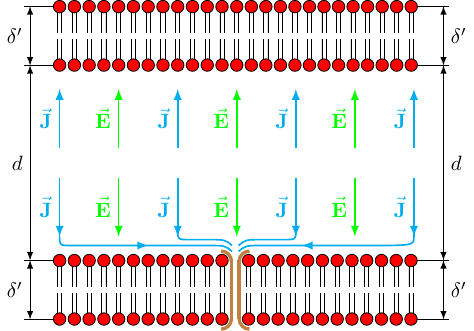}
\vskip  10pt
\includegraphics[width=0.83\fw]{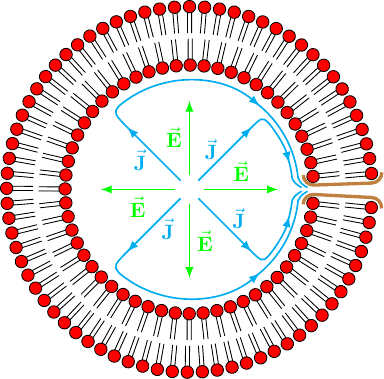}
\caption[]{Schematic picture of an axial (top) and a radial (bottom)
  sections of the axon. A radial component of an electric field
  $\vec{E}$ inside the axon produces only a radial component of a
  current $\vec{J}$ with typical isotropic conductivity. But a
  presence of cross-membrane channels results in appearance of non
  isotropic dependence of current as a response to a supplied electric
  field, giving rise to axial (top) and azimuthal
  (bottom) components of the current.}
\label{fig:axon}
\end{figure}

A simplified schematic picture of this anisotropic electric field --
electric current geometry is shown in \cref{fig:axon}, although it is
shown not to scale as $\dl \ll 1$ and all the anisotropic currents
should be shown in the very thin boundary layer and not far outside of
the $1\le r\le 1+\dl$ ring.  Nevertheless, the schematics can be
useful to emphasize the highly anisotropic structure of the
voltage-current relationship when the membrane interface is present.

\subsection{Solutions to the field equations}

The solution to the charge continuity equation {\Cref{eq:phiTensor}}
within this anisotropic and inhomogeneous axon geometry is all that is
required to explain the generation of surface waves that propagate
through the extracellular-intracellular membrane interface.  We
emphasize that this is a derivation from first physical principles, in
contradistinction to the standard model constructed from multiple
empirical equations with multiple empirically fitted constants
\citep{pmid12991237}. To simplify the math in a similar fashion as our
previously published work {\citep{Galinsky:2020a}} and provide a more
intuitively clear result, we will assume the axon to be described by
an axially symmetric cylindrical geometry \citep{Scott1975-ht},
although, generally speaking, for $\dl \ll 1$ this is not absolutely
necessary. 

Defining
\begin{align}
\label{eq:uSv}
\mvec{u} \equiv
\begin{pmatrix}
\frac{1}{r}\D{}{r}r \\
\\
\D{}{z}
\end{pmatrix} \sscomma
\mvec{\Sigma} \equiv
\begin{pmatrix}
\Sigma_{rr} & \Sigma_{rz}\\
\Sigma_{zr} & \Sigma_{zz}
\end{pmatrix} \sscomma
\mvec{v} \equiv
\begin{pmatrix}
\D{\phi}{r} \\
\\
\D{\phi}{z}
\end{pmatrix}
\end{align}
equation \cref{eq:phiSigma} can be
written for a single axon in cylindrical $(r,z)$ coordinate system as
\begin{align}
\label{eq:phiRZ}
&\D{}{t} \paren{\mvec{u}^{t} \cdot \mvec{v}}
= -\mvec{u}^{t} \cdot \mvec{\Sigma} \cdot \mvec{v} 
\end{align}
Because of the huge difference in scaled conductances inside and
outside of the bilayer membrane, the assumption of perfectly
conducting boundary condition on both sides of the membrane bilayer is
accurate and explicit solutions for the extracellular and the
intracellular space are not required. It is only necessary to solve
equation \cref{eq:phiRZ} inside the ring $1\le r \le 1+\dl$. We will
seek the solution in the form
\begin{align}
\label{eq:phiM}
\phi(r,z,t) &= \phi^{0}(r) + \phi^\prime(r,z,t),\\
\label{eq:phiEq}
\phi^{0}(r) &= \frac{\ln r}{\ln (1+\dl)} \approx
\frac{1}{\dl}\ln r,
\end{align}
where $\phi^{0}(r)$ is a stationary, time independent (or equilibrium)
solution of the equation \cref{eq:phiRZ} inside the ring $1\le r \le
1+\dl$, such that $\phi^{0}(r) \le 1$ anywhere inside the ring whereas outside the ring
$\phi^{0}(r) = 0$ for $r\le 1$ and $\phi^{0}(r) = 1$ for $r \ge 1+\dl$.

The solution to the field equations can be approached at two levels of
accuracy, a simplified but intuitive linear version, and a more
accurate but complex non-linear version, by formally expanding the
nonlinear dependence of the conductivity tensor in dimensionless form
into Taylor series as
\begin{align}
\label{eq:SigmaT}
&\bm{\Sigma}(\phi) = \bm{\Sigma}^0 + \frac{V}{V_0} \bm{\Sigma}^\prime\phi + \dots,\\
\mbox{where}\quad
&\bm{\Sigma}^0\equiv \bm{\Sigma}(\phi)|_{\phi=0}\quad \mbox{and}\quad
\bm{\Sigma}^\prime \equiv
\left.\D{\bm{\Sigma}(\phi)}{\phi}\right|_{\phi=0}  
\end{align}
where {\cref{eq:SigmaT}} has been constructed
with an adjustable normalization $V$ to facilitate the inclusion of
external conditions such as those prevalent in a wide range of
experiments.  For example, it can be set to the equilibrium value of
a voltage drop across the membrane for voltage clamping
experiments.

Without a loss of generality we can assume that the zeroth order terms
are axisymmetric, with average dimensionless cross membrane
conductivity $0 < \Sigma^0_{rr} \equiv \Sigma^0 \ll 1 $, average
conductivity along the membrane (that possibly is significantly
smaller) $\Sigma^0_{zz} = \epsilon\Sigma^0_{rr}$ ($\epsilon < 1$), and
zero off-diagonal terms $\Sigma^0_{rz} = \Sigma^0_{zr} = 0$, i.e.,
\begin{align}
\label{eq:Sigma0}
\bm{\Sigma}^0 &=\Sigma^0
\begin{pmatrix}
1 & 0\\
0 & \epsilon
\end{pmatrix}.
\end{align}

With this positive definite matrix form used for $\bm{\Sigma}^0$ the
only solution that can be obtained from the equation \cref{eq:phiRZ}
will correspond to the loss of the electrostatic field energy in the
membrane. In order to be able to compensate for this loss and to keep
the potential difference across the membrane at a fixed ``resting
potential'' level some additional mechanisms are required. In axons
this happens by adding energy through ATP-mediated diffusion. For the
purpose of this paper, we are not interested in the details of this
process and we will just assume that it provides required amount of
energy to keep the cross membrane voltage drop at a constant level.

Because of the different concentrations of the different ions in
extracellular and intracellular fluids (in particular, sodium and
potassium ions), it has been known for a long time
that nonlinear membrane properties show a positive feedback effect for
the radial current-voltage relationship \citep{Scott1975-ht}. In terms
of the nonlinear passive response produced by the conductivity tensor
it means that some of the $\bm{\Sigma}^\prime$ components are
negative. At the same time the structure of $\bm{\Sigma}^\prime$
should guarantee that there is neither total (volume integrated)
additional electrostatic energy loss nor total electrostatic energy
generation produced due to this nonlinear self coupling, therefore
both eigenvalues of $\bm{\Sigma}^\prime$ should be zeros
(the eigenvalues of the conductivity matrix are real).
As membrane
conductivity is normalized by $\Sigma^i$, we would require that
$|\Sigma_{\{\dots\}}| \le 1$, and will assume that both
$|\Sigma^0_{\{\dots\}}|$ and $|\Sigma^\prime_{\{\dots\}}|$ are less
than 1.  This limits the structure of $\bm{\Sigma}^\prime$ to the
following form
\begin{align}
\label{eq:Sigma1}
{
\bm{\Sigma}^\prime =\Sigma^\prime
\begin{pmatrix}
s_\perp xy & -s_\parallel x^2 \\
s_\parallel y^2 & -s_\perp xy
\end{pmatrix},}
\end{align}
where $\Sigma^\prime = \max{\abs{\Sigma^{\prime}_{ij}}}$,
$\max(x,y)=1$, $\min(x,y)\ge 0$, and
both $s_\perp$ and $s_\parallel$
can either be -1 or 1. As we will see below, the choice of
$s_\parallel$ between -1 and 1 is not particular important, as it
simply selects different directions of wave propagation, but the
different choice for a sign of $s_\perp$ selects different scales
where wave excitation and/or damping occurs, that experimentally has
been noted as a different behavior of spiking for Type I and Type II
neurons.

Based on experimental results that we cited above
\citep{Scott1975-ht,Bedard2004-lz}, the normalized linear membrane
conductivity is expected to be significantly less than 1
($|\Sigma^0_{\{\dots\}}| \sim 10^{-8} - 10^{-12} \ll 1$). Therefore,
the assumption for the first order normalized membrane conductivity
that $|\Sigma^\prime_{\{\dots\}}|<1$ does not require it to be smaller
than the linear normalized membrane conductivity, on the contrary it
may be expected that $1 > |\Sigma^\prime_{\{\dots\}}| \gg
|\Sigma^0_{\{\dots\}}|$.

The solution for the second term $\phi^\prime(r,z,t)$ in equation
\cref{eq:phiM} can be expanded using radial and axial eigenmodes of
the linearized system, with perfectly conducting boundary conditions
at $r=1$ and $r=1+\dl$ that require that $E_z=0$ or
$\phi_{r}(1)=\phi_{r}(1+\dl) = 0$.
\begin{align}
\label{eq:phiEV}
\phi_{r}(r) \sim R_0(\kappa r + \eta), \quad \phi_{w}(z,t) \sim e^{-i
  (\omega_k t + k z)}, 
\end{align}
where $R_0$ denotes Bessel functions either of the first ($J_0$) or
the second ($Y_0$) kind, and $\kappa$ and $\eta$ can be determined
from the boundary conditions, $R_0(\kappa + \eta)=R_0(\kappa(1+\dl)
+\eta)=0$. Note that the parameters $\kappa$ in $\phi_{r}(r)$ plays a
similar role as the axial wave number $k$ in $\phi_{w}(z,t)$ as larger
values produce shorter wavelength spatial oscillations, but in the
radial direction (although we are not interested in different radial
modes and simply assume an existence of the longest mode with
$\kappa\dl \sim 1$). The derivative of radial eigenmode can then be
written
\begin{align}
\label{eq:phiDer}
\der{\phi_{r}}{r} \sim -\kappa R_1(\kappa r + \eta),
\end{align}
where again $R_{1}$ denotes Bessel functions either of the first
($J_{1}$) or the second ($Y_{1}$) kind, and $R_1(\kappa + \eta)
\approx \pm R_1(\kappa(1+\dl) + \eta)$.

Proceeding in a spirit similar to our earlier analysis
\cite{Galinsky:2020a}, we first solve the simpler linear wave
analysis problem by considering only the linear terms in equation
{\cref{eq:SigmaT}} which are independent of $z$ and $t$, then expand
the scope of the analysis to include the non-linear terms that
depend on $z$ and $t$.

\subsection{Linear wave analysis and surface wave generation}
The linear in $\phi^\prime(r,z,t)$
terms in equation \cref{eq:phiRZ} that are independent of
$z$ and $t$ include from equation \cref{eq:SigmaT}
$\Sigma^0_{\{\dots\}}$, that are constant inside the membrane layer,
and $\Sigma^\prime_{\{\dots\}} \phi^0(r)$, that only depend on
radius $r$. Substituting the eigenmode solutions
\cref{eq:phiEV} into \cref{eq:phiRZ}, multiplying by $\phi_{r}(r) r$,
and integrating the radial part across the membrane bilayer, we obtain
the complex dispersion relation
\begin{align}
\label{eq:dispersion}
i \Omega_k &\equiv \gamma_k + i \omega_k = \Lambda_\perp  + i \Lambda_{\parallel} k
\end{align}
and the real $\Lambda_\perp$ and the imaginary $i\Lambda_{\parallel}k$
parts of the dispersion correspond to the diagonal and the off
diagonal conductivity tensor components,
\begin{align}
\label{eq:Sigpp}
\gamma_k &\equiv \Lambda_\perp = \left(\gamma_d - \gamma_e\right),\\
\label{eq:gamma_d}
\gamma_d &= \frac{\Sigma^0}{\varkappa^2} \paren{\kappa^2+\epsilon k^2}, \\
\label{eq:gamma_e}
\gamma_e &= \hat{V}  \Sigma^\prime \frac{s_\perp x y}{\varkappa^2}
 \frac{\paren{\kappa^2 C_\perp^r-k^2  C_\perp^z}}{\dl C} 
\nn & 
\approx
\hat{V} \Sigma^\prime \frac{s_\perp x y }{2\varkappa^2} \paren{\kappa^2-k^2},\\
\label{eq:lam_para1}
\Lambda_\parallel  &= \hat{V}  \Sigma^\prime \frac{C_\parallel}{2 \dl C}
\frac{s_\parallel (x^2+ y^2)}{\varkappa^2} 
\approx 
\hat{V}  \Sigma^\prime \frac{s_\parallel (x^2 + y^2)}{2 \dl \varkappa^2},
\end{align}
in which $\hat{V} \equiv V/V_{0}$ is the fractional voltage (i.e., the
fraction of the resting potential occupied by the external voltage) and
\begin{align}
\label{eq:varkappa}
\varkappa^2 &\equiv \kappa^2+k^2 
\end{align}
The normalization parameters $C_\perp$, $C_\parallel$, and $C$ are
provided in \Cref{app:norm} by \cref{eq:phiInt}.  The parameter
$\varkappa^2$ can be viewed as the length (squared) of a vector
$\bm{\varkappa} = \bm{\kappa} + \bm{k}$ in an abstract vector space
that controls the spatial scale of oscillations in the radial and
longitudinal (axial) coordinates of the axon. The component
$\Lambda_\perp$ describes the damping ($\gamma_d$) or excitation
($\gamma_e$) of the waves while $\Lambda_\parallel$ is related to the
wave oscillations $\omega_{k}$.  Expressions
\cmmnt{\Cref{eq:lam_para1}} \cref{eq:gamma_d} and
\cref{eq:lam_para1} can be approximated as \cmmnt{(\Cref{app:disp})}
\begin{align}
\label{eq:lam_perp}
\Lambda_{\perp} &\approx
\Sigma^0 \lims{
\paren{\hat{\kappa}^2 + \epsilon \hat{k}^2}
+ 
\hat{\sigma}_{\perp} \paren{\hat{k}^2 - \hat{\kappa}^2}
}
\\
\label{eq:lam_para}
\Lambda_\parallel &\approx
\Sigma^0 \frac{\hat{\sigma}_{\parallel}}{\dl \varkappa^2},
\end{align}
where $\hat{\kappa} \equiv \kappa/\varkappa$ and $\hat{k} \equiv
k/\varkappa$ are the fractional wave numbers and
\begin{align}
\label{eq:sig_perp}
\hat{\sigma}_{\perp} &\equiv \frac{1}{2} \hat{V} \hat{\Sigma} \, s_{\perp}xy \\
\label{eq:sig_para}
\hat{\sigma}_{\parallel} &\equiv \frac{1}{2} \hat{V} \hat{\Sigma}\, s_\parallel (x^2 + y^2) 
\end{align}
where $\hat{\Sigma} \equiv \Sigma'/\Sigma_{0}$ is the fractional
conductivity (i.e., the ratio of the conductivity perturbation
magnitude to the mean membrane conductivity).
The parameters
$\hat{\sigma}_{\perp}$ and $\hat{\sigma}_{\parallel}$ are the
weightings for the (fractional) radial and longitudinal wave vector
contributions to the radial and parallel components, respectively, of
the dispersion relation.  Each is scaled by both the fractional
voltage and the fractional conductivity. The radial and
longitudinal are scaled, respectively, by $s_{\perp} = \pm 1$ and
$s_{\parallel} = \pm 1$ that have been introduced to demonstrate the
profoundly different wave characteristics possible within the
available parameter ranges of \cref{eq:dispersion}.

\subsubsection{The existence of waves}
This solution to the simplified linear problem is sufficient to show a
key result - the existence of propagating surface waves along the
axon.  To see this, note that for large $k$ $(k\gg\kappa)$ that
$\varkappa \approx k$ so from \cref{eq:lam_para},
$\Lambda_\parallel \sim 1/k^2$ so that the oscillatory
component of the dispersion relation {\cref{eq:dispersion}} is
approximately $\omega_k =\Lambda_{\parallel} k \sim 1/k$ and thus
exhibits the same inverse proportionality of frequency and wave number
shown in our previous work \cite{Galinsky:2020a,Galinsky:2020b} (using
Cartesian geometry) to generate surface (or interface) electric field
waves.  The relative magnitude of the conductivity tensor
components in \cref{eq:SigmaT} are such that $1 >
|\Sigma^\prime_{\{\dots\}}| \gg |\Sigma^0_{\{\dots\}}|$ so that
$\hat{\Sigma} \gg 1$ and thus the fractional conductivities
\cref{eq:sig_perp} and \cref{eq:sig_para} provide sufficiently large
parameter ranges within the membrane to support wave excitation.

\subsubsection{Wave characteristics}

The parameters $s_\parallel = \pm 1$ and $s_\perp = \pm 1$
were introduced to delineate the profoundly different parameter
regions of the dispersion relation \cref{eq:dispersion}. This can
now be shown directly using \cref{eq:lam_para} and
\cref{eq:lam_perp}.

First consider the parallel component $s_\parallel$.  Note
that phase velocity of the waves is defined as $\omega_k/k =
\Lambda_\parallel$ and thus determined by \cref{eq:lam_para}.
Changing the sign of the $s_\parallel$ changes the sign of
$\sigma_\parallel$ in \cref{eq:sig_para} and thus changes the sign
of the phase velocity \cref{eq:lam_para}.  That is, it changes the
direction of the wave propagation. 

Now consider the perpendicular component $s_\perp$. Its
influence on the solution provides an important new understanding of
the role of the Nodes of Ranvier.  Changing sign of $s_\perp$ will
change the sign of the wave excitation rate $\gamma_e$
\cref{eq:gamma_e}, i.e., it will result in two different wave
excitation patterns. If $s_\perp$ is positive, $\gamma_e>0$ when
$k<\kappa$, i.e., waves with longer wave length will be excited, which
corresponds to Type I myelinated axons, where longer wave lengths are
preset by the internodal distances between the Nodes of Ranvier and
the maximum wave length will be determined by the strongest excitation
at the internodal length. 
For $s_\perp=-1$ the wave
excitation rate $\gamma_e$ will be positive for $k> \kappa$ and will
be increasing with the increase of the wave number $k$, hence shorter
scale waves (often at the subthreshold level) and higher frequencies
will be seen, i.e., more representative of unmyelinated Type II (and
possibly some unmyelinated Type I as well) behavior. 

\subsection{Wave speeds and myelination}
The dispersion relation allows the calculation of the wave
phase velocity, the rate at which a wave of a single frequency
propagates through the medium.  The dimensional wave phase speed $\mathcal{V}$
for the component along the axon from
{\cref{eq:dispersion}} is
\begin{align}
\label{eq:Vdef}
\mathcal{V} \equiv \frac{\omega_{k}}{k} \Sigma^{i} d =
{\Lambda}_{\parallel}\Sigma^{i} d
\end{align}
where the factor
$\Sigma^{i} d$ appeared as the parameters have been converted to dimensional form. 
The simple estimates of wave phase velocity, in particular the
dependence of the velocity on axon diameter $d$, show consistent
behavior with both myelinated and unmyelinated
conditions. 

\paragraph{Myelinated axons}
For myelinated axons the ratio of the axon diameter to the total (axon
and myelin) diameter is relatively constant (around 0.6-0.8)
\citep{Gillespie1983-qd,Arancibia-Carcamo2017-pe} so that in our
dimensionless units $\dl_m\sim 0.2-0.4$. This determines the radial
oscillation spatial wave number $\kappa_m\sim \pi/\dl_m\sim 5 - 15$.
As myelination makes the cross membrane conductivity ($\Sigma^0_{rr}$)
smaller, it effectively decreases the wave damping $\gamma_d$ for all
scales smaller than the inter-node distance. Therefore, we may assume
that the inter-node distance between the Nodes of Ranvier $L_m$
determines the wavelength of the propagating modes.  The inter-node
distance between Nodes of Ranvier $L_m$ can be as high as 1.5 mm, but
typically ranges from 350 $\mu$m for 12 $\mu$m axon diameter, to 205
$\mu$m for 3.4 $\mu$m axon diameter \citep{Waxman1971-fm}, to 139
$\mu$m for 0.82 $\mu$m axon diameter \citep{Arancibia-Carcamo2017-pe}
so that parallel spatial wave number $k_m\sim2\pi/L_m \sim 0.005 -
0.05$ and $\kappa \gg k_{m}$ so that $\varkappa \approx \kappa$.
Hence, for myelinated axons the wave phase speed is directly
proportional to axonal diameter (assuming that
$\Sigma^\prime_\parallel = (x^2 + y^2) \hat{V}\Sigma^\prime/2 \sim
0.05$, i.e., less than a maximum value of 1 due to multiple layers of
myelin, $d$ is in the units of $\mu$m, and a conversion factor from
$\mu$m to $m$ is included into the numerical constant)
\begin{align}
\label{eq:myelinV}
\mathcal{V} = \Lambda_\parallel\Sigma^i d \sim 5 \times 10^{3}
\frac{\Sigma^\prime_\parallel}{\dl_m \kappa_m^2}
d \sim (5-10)d\,
\end{align}
in units of $m/s$, giving values of 100-200 m/s for 20
$\mu$m diameter axons which is consistent with published values
\citep{Siegel2005-dh}.

These results also provide an explanation for some recently
detected anomalous phenomena of nerve conduction, 
such as the observations that in myelinated nerves the
conduction velocity increases with stretch which
contradicts existing theories since the diameter decreases on
stretching \cite{Sharmin2023-sf}.  However,
this agrees well with our results as stretching increases the
intra-nodal distance, hence increases both the wave length and the
wave phase velocity \cref{eq:lam_para}.

\paragraph{Unmyelinated axons}
For unmyelinated axons the membrane diameter is constant
$\delta_u\sim$ 10 nm = $10^{-2} \mu$m and the wavelength $L_u$ of the
propagating modes is going to be significantly smaller (depending on
the small scale membrane geometry), but it is reasonable to assume
$L_u\sim d/10$. That gives for the dimensionless wavenumber $k_u \sim
2\pi d/L_u \sim 20\pi\sim 10^2$, and $\kappa_u$ is again determined by
the same relation $\kappa_u\sim \pi/\dl_u$, where $\dl_u$ now is not
fixed, $\dl_u=\delta_u/d$. Then the expression to the wave speed as a
function of $d$ (assuming maximum value for 
$\Sigma^\prime_\parallel\sim 1$, and both $d$ and 
$\delta_u$ in the units of
$\mu$m)
\begin{align}
\label{eq:nonmyelinV}
\mathcal{V} = \Lambda_\parallel\Sigma^i d \sim 5 \times 10^{3}
\frac{\Sigma^\prime_\parallel}{\delta_u
  k_u^2}\frac{d^2}{1+d^2\pi^2/(\delta_u^2k_u^2)}, 
\end{align}
giving roughly range from 0.5 to 5 m/s for axon diameters from 0.1 to
10 $\mu$m Thus the wave speeds of myelinated axons are
predicted to be around two orders of magnitude larger than
unmyelinated axons.  The importance of this analysis is not just
that these predictions are consistent with measured values but that
they were derived from first principles and therefore based on
rather simple (at least to first order) measurable axon
characteristics.  This offers the potential for a better
understanding of brain communication deficits associated with
ubiquitous demyelinating diseases such as multiple sclerosis.

\subsection{Diffusion limits of the cable theory}
\label{sec:cable}

A standard accepted model for the propagation of the action potential
spike is the so-called \textit{cable theory}, an approach developed by
Hodgkin and Rushton \citep{Hodgkin:1946} to model the passive
conduction based on theoretical work on submarine telegraph cables by
William Thompson (Lord Kelvin).  This work was further developed and
extended to dendritic spines by Rall \citep{Rall:1962,Rall2011-hr} who
popularized this approach which has now become an established model in
description of neuronal communication
\citep{Segev:2000,Holmes2014-pl,Pagkalos2023,Spruston:2013}
with a host of variations, such as double cables
\cite{Cohen2020-ie, Lim2020-sq}.  Because of the linear cable
theory's ubiquity and universal acceptance, we take a brief digression
in this section to demonstrate that it is derivable from our more
general theory described by \cref{eq:phiRZ} by making several
simplifications.  In doing so, we reveal that the standard cable
theory does not actually support sustained propagation of the action
potential in a wide range of experimentally reported physiological
parameters.

\paragraph{Derivation of the cable equation from \cref{eq:phiRZ}}
The importance of tissue anisotropy and inhomogeneity and a full
non-linear analysis to the generation of persistent surface waves is
emphasized by the fact that the cable equation, which does
\textit{not} produce such waves, can be recovered from our general
model \cref{eq:phiRZ} by ignoring important components that contribute
to these properties, namely the non-diagonal and non-linear terms in
the conductivity tensors.

Ignoring all non-diagonal and nonlinear terms in the conductivity
tensors and assuming that only $\Sigma^0_\perp$ and
$\Sigma^0_\parallel$ terms are non-zero,
so that $\mvec{\Sigma}$ in \cref{eq:uSv}
\begin{align}
\mvec{\Sigma} \equiv
\begin{pmatrix}
\Sigma^{0}_{\perp} & 0\\
0 & \Sigma^{0}_{\parallel}
\end{pmatrix} \sscomma
\end{align}
the equation for the electric
field potential becomes, from \cref{eq:phiRZ},
\begin{align}
\label{eq:precable}
&\hspace{-5pt}\D{}{t} \left(\frac{1}{r}\D{}{r}r\D{\phi}{r} + \D{^2\phi}{z^2}\right) =
-\Sigma^0_\perp\frac{1}{r}\D{}{r}r\D{\phi}{r}
-\Sigma^0_\parallel\D{^2\phi}{z^2}.
\end{align}
As cable equation is not supposed to follow the exact radial
dependence of the $\phi$, we can use the above equation
\cref{eq:precable} and obtain its approximate form in the limit of a
very thin lipid bilayer, i.e., $\dl \ll 1$ and assuming that the
largest radial variations of the potential $\phi$ are located around
the membrane. This enables an approximate solution where the time
dependence of the field is wholly contained in the axial dimension,
while the radial component is constructed to meet some minimal
boundary conditions based on simple geometric constraints. Therefore,
we can search for the approximate solution separable in the radial and
axial dimensions of the form $\phi=\phi^\prime_r(r)\phi_{a}(z,t)$,
where $\phi^\prime_r(r)=-1+\phi^0(r)$, i.e., $-1$ for $0\le r \le 1$,
transitions from -1 to 0 for $1 \le r\le1+\dl$ and equals 0 for
$r>1+\dl$ . Multiplying the equation \cref{eq:precable} by
$\phi^\prime_r(r) r$ and integrating it from 0 to infinity we obtain
the cable equation in the usual form \citep{Rall2011-hr,Holmes2014-pl}
as
\begin{align}
\label{eq:cable}
&\hspace{-5pt}\frac{1}{\dl}\D{\phi_{a}}{t}
+\frac{\Sigma^0_\perp}{\dl}\phi_{a} =
\frac{\Sigma^0_\parallel}{2}\D{^2\phi_{a}}{z^2},
\end{align}
where we used 
\begin{align}
&\!\!\!\!\int\limits_0^\infty
\frac{1}{r}\D{}{r}r\D{\phi}{r} \phi^\prime_r(r) r dr = -
\phi_{a}\!\!\int\limits_1^{1+\dl}\left|\D{\phi^\prime_r}{r}\right|^2 r dr 
\approx -\frac{\phi_{a}}{\dl},\\
&\!\!\!\!\int\limits_0^\infty
\phi \phi^\prime_r(r) r dr =
\phi_{a}\int\limits_0^{1+\dl}|\phi^\prime_r(r)|^2 r dr \approx
\frac{\phi_{a}}{2}. 
\end{align}
and have ignored the second term under time derivative in
{\cref{eq:precable}} because it is negligible when the axial scales of
variation of the potential $\phi_{a}$ is larger than $\sqrt{\dl/2}$.
We thus recover the cable equation \cref{eq:cable} from
\cref{eq:precable}, where $\Sigma^0_\perp$ corresponds to the
normalized membrane scaled conductivity and $\Sigma^0_\parallel$
equals to the normalized scaled conductivity of the axon internal
fluid (i.e., $\Sigma^0_\parallel \sim$ 1 in dimensionless units). The
terms in the equation \cref{eq:cable} directly correspond to
dissipative (no positive feedback) terms of $\Lambda_\perp$
\cref{eq:Sigpp,eq:gamma_d,eq:gamma_e} in the limit $k^2\ll\kappa^2
\sim 1/\dl^2$.

\paragraph{Length and time scale of the cable equation}

The cable equation describes that the height of the action potential
peak decays with time $t$ as $\sqrt{t_0/t}$, where the shortest time
$t_0 \sim 1/\Sigma^0_\parallel$ corresponds to the narrowest
($\sqrt{\dl/2}$) and the tallest ($\phi_{a}(t_0,z_0) = \phi_m $) shape
of the action potential peak that the cable equation is capable of
describing (and when the cable equation approximation is valid). The
decay is actually even faster as it includes an exponential term
$\exp(-\/\Sigma^0_\perp t)$, but the approximate time dependence will
be valid when $t<1/\Sigma^0_\perp$ and, as the ratio of the cross
membrane to the intracellular conductivities is very small ($10^{-8}$
to $10^{-12}$) \citep{Bedard2004-lz}, we can safely use this
approximation in all our estimates below.  The dimensionless diffusion
coefficient is then equal to $\Sigma^0_\parallel \dl /2$,
which allows us to find the time
dependence for the axial diffusion length (the half width of the
pulse) as $\Delta z \sim \sqrt{t \Sigma^0_\parallel \dl /2}$.

The ratio of the differences of the action potential firing threshold
to the total peak above the resting potential in the HH-model is equal
to about $\Delta\phi_{a}/\phi_{m} \sim 0.15$ (resting potential is -70
mV, threshold -55 mV, peak 30 mV).  Therefore, the maximum time until
the diffusively spreading pulse will be above threshold is
about $\sim t_0 \phi_m^2/{\Delta\phi_{a}}^2\ll1/\Sigma^0_\perp$,
giving for the maximum diffusion length $\Delta z \sim
\sqrt{\dl/2}\phi_m/\Delta\phi_{a}$.

\paragraph{Myelinated axons} For a myelinated 20 $\mu$m
diameter axon with the thickest myelin layer ($\dl \sim 0.4$) this
gives the maximum diffusion length of about 60 $\mu$m, which is
significantly shorter than the internodal length of $\sim 2$mm between
the Nodes of Ranvier of the typical 20 $\mu$m axon.  Decreasing the
myelin thickness will decrease the maximum propagation length even
more.  Na\"ive attempts to adjust the threshold parameters of HH-model
to accommodate for longer maximum diffusion length will quickly reveal
the significant model inconsistencies. For example, using the typical
average ranges of internodal distances for different axon diameters
(350 $\mu$m for 12 $\mu$m axon diameter, 205 $\mu$m for 3.4 $\mu$m
axon diameter \citep{Waxman1971-fm}, to 139 $\mu$m for 0.82 $\mu$m
axon diameter \citep{Arancibia-Carcamo2017-pe}) it can be easily seen
that it will require to decrease the firing threshold in 10 to 60
times ($350 \times 0.15/12/\sqrt{\dl/2} \sim 10$, $205\times
0.15/3.4/\sqrt{\dl/2} \sim 20$, $139\times 0.15/0.82/\sqrt{\dl/2} \sim
60$), hence will require exceedingly (and unrealistically) low
threshold voltages in the range of -68.5 to -69.75 mV instead of -55
mV for the resting potential of -70 mV.  The conclusion is therefore
that the amount of diffusion provided by standard cable theory to
action potential spike generation by thresholded reactive HH mechanism
with experimentally confirmed parameter values is generally
incompatible with the process of saltatory conduction.

\paragraph{Unmyelinated axons} For unmyelinated axon the original HH
model assumes that there are 60 Na$^+$ channels and 18 K$^+$ channels
for every $\mu$m$^2$ of membrane \citep{Sengupta2013-ya}.  A more
detailed analyses of variations of Na$^+$ channels density show that
in unmyelinated hippocampal axons, the density increases tenfold from
the soma with 2.6 channels/$\mu$m$^2$, through the proximal axon (25
channels/$\mu$m$^2$), to the distal axon (46.1 channels/$\mu$m$^2$)
\citep{Hu2014-lj,Freeman2016-ph}. Therefore, it can be safely assumed
that the average linear distances between ion channels for
unmyelinated axon change from about 0.1 $\mu$m to 0.7 $\mu$m ($\sim
1/\sqrt{2.6}$).  For a 500 $\mu$m diameter giant squid axon with 10 nm
membrane, like the one used by Hodgkin and Huxley in their seminal
work, the maximum
diffusion length is about $\sqrt{10^{-2}/2/500}\times 500/0.15 \sim
10.5 \mu$m, which is significantly above the average linear
inter-channel distance of 0.1 $\mu$m, thus gives enough flexibility to
successfully do a mind entertaining exercise of fitting diffusive and
reactive processes together.  But for thin unmyelinated pyramidal
tract dendrites with around 0.2 $\mu$m diameter and 5 nm membrane
thickness ($\dl \sim 0.025$), the maximum propagation distance is
about 0.15 $\mu$m, i.e., several times less than 5 Na$^+$
channels/$\mu$m$^2$ and 5 K$^+$ channels/$\mu$m$^2$ density of
pyramidal neuron \citep{Arhem2006-dl} would provide.

The conclusion is that the action potential propagation model
described by cable theory is incompatible with experimentally measured
physiological parameters of both myelinated and unmyelinated axons.
On the contrary, our linear wave model developed from first principles
using measured physiological tissue parameters is able to describe
wave propagation in all these parameter ranges.

\subsection{Non-linear wave analysis}

The linear wave analysis above is sufficient to demonstrate the
existence of sustained propagating waves along the axons.  However, as
demonstrated in our previous work
\citep{Galinsky:2020a,Galinsky:2020b}, a full non-linear analysis is
necessary to accurately describe the details of the spatiotemporal
characteristics of the propagating waves.

Proceeding as in \citep{Galinsky:2020a,Galinsky:2020b}, the solution
$\phi_{w}(z,t)$ is expanded using a Fourier integral

\begin{align}
\label{eq:phiW}
\phi_{w}(z,t) = \int\limits_{-\infty}^{\infty}
a_k(t) e^{i\left(k z + \omega_{k}t\right)}dk + c.c.,
\end{align}
assuming that 
\begin{align}
\left|\frac{1}{a_k(t)}\der{a_k(t)}{t}\right| < \omega_k.
\end{align}
and where c.c.~refers to the complex conjugate.
This results in a set of 
coupled equations for time dependent complex
amplitudes $a_k(t)\equiv a(k,t)$
\begin{align}
\label{eq:phiAk}
\der{a_k}{t} &= (\gamma_e - \gamma_d) a_k + \mathcal{N}_k,
\end{align}
that have the same general form as Equation 14
in \citep{Galinsky:2020a}, where
\begin{align}
\label{eq:Nphi}
\mathcal{N}(\phi) &=  D_\perp \phi_{w}^2 +
D\der{\phi_{w}^2}{z} +  
D_\parallel \der{^2\phi_{w}^2}{z^2},\\
\mathcal{N}_k &= \frac{1}{2\pi}\int\limits_{-\infty}^{\infty}
\mathcal{N}(\phi) e^{-i\left(k z + \omega_{k}t\right)}dz,
\end{align}
where the normalization coefficients $D_\perp$, $D_\parallel$, and $D$
are given in \Cref{app:coef} by \cref{eq:D}.  The detailed evaluation
of nonlinear input from multiple wave modes assuming a general
quadratic form of nonlinearity was shown in detail for both
non-resonant and resonant terms in
\citep{Galinsky:2020a,Galinsky:2020b}. It was shown there for the
first time that it is the inverse proportionality between frequencies
and wave modes that allows calculation of the nonlinear input in a
relatively simple analytical form, resulting in a simple nonlinear
equation for wave amplitude $a_k(t)$. Following
\citep{Galinsky:2021b,Galinsky:2023a,Galinsky:2023b} this equation can
be written in the general form
\begin{align}
\label{eq:ak}
\der{a_k}{t}&=\gamma_k a_k + \beta^\prime_{k} a_k a_k^* + \beta_k a_k^2
- \alpha_k a_k (a_k a_k^* )^{1/2},
\end{align}
where complex $\gamma_k$ includes $\gamma_e-\gamma_d$ as a real part
and $\omega_k$ as an imaginary part, and the parameters $\alpha$,
$\beta$, and $\beta^\prime$ can be evaluated following
\citep{Galinsky:2020a,Galinsky:2020b} using coefficients in
\cref{eq:phiAk,eq:Nphi}.

\begin{figure}[!tbh] 
\centering
\includegraphics[width=\fw]{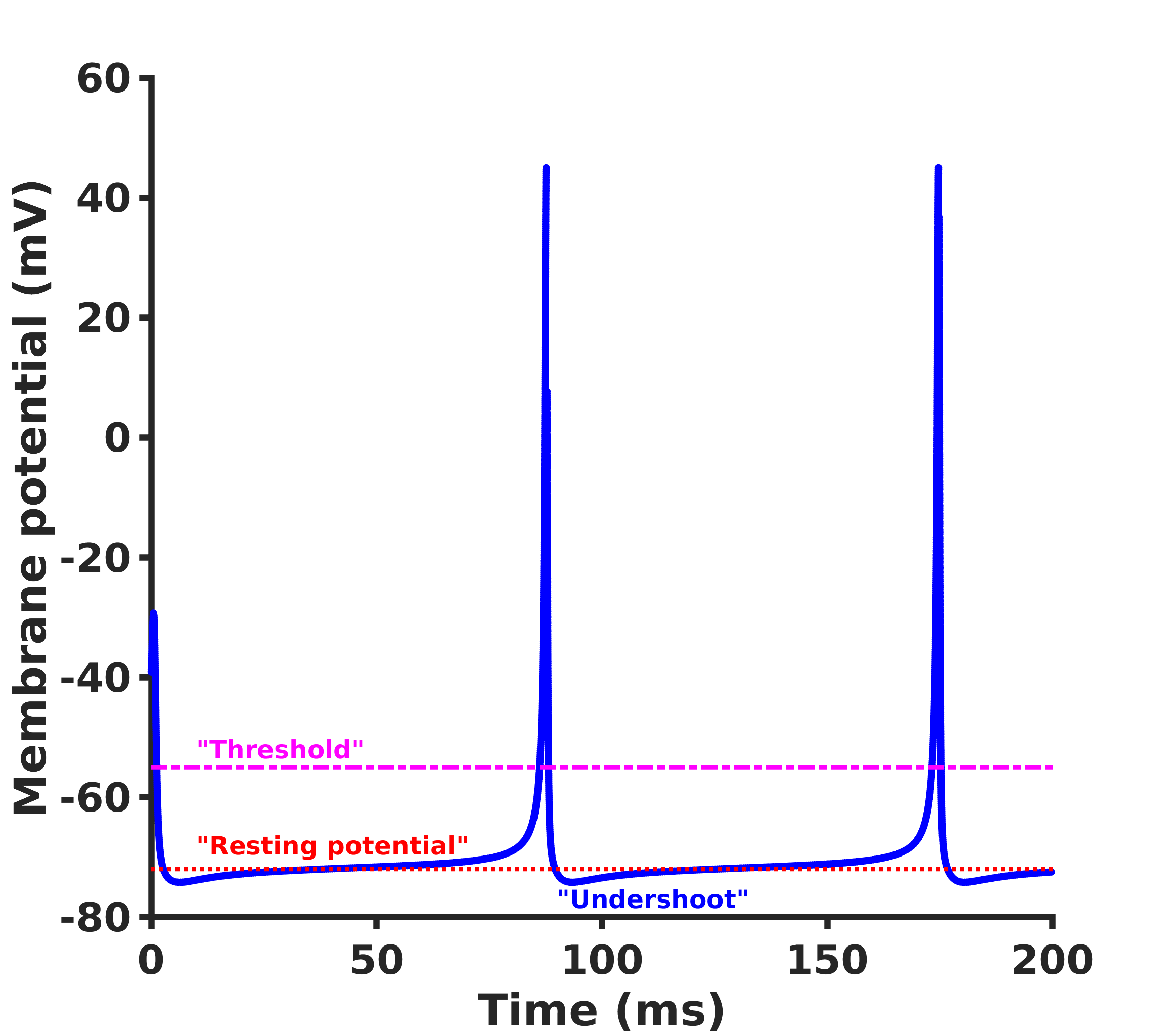}
\vskip  10pt
\includegraphics[width=\fw]{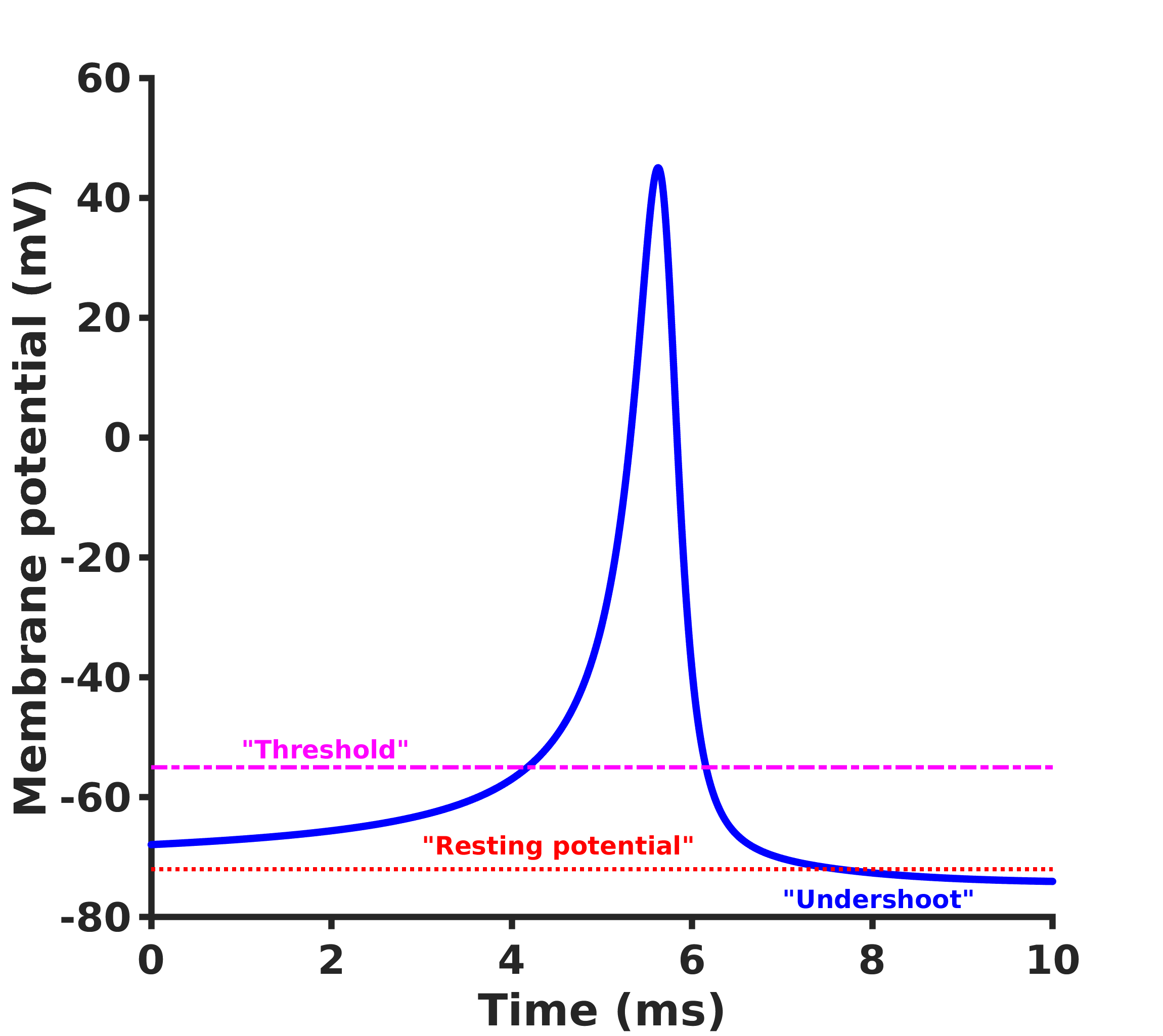}
\caption[]{An example of a numerical solution of equation \cref{eq:ak}
  (top) and an expanded view of a single spike (bottom) using
  $\beta_k^\prime = \exp(i\pi/4)$, $\beta_k = 2\exp(-i\pi/4)$,
  $\alpha_k = 3$, and $\gamma_k = 1.996 + i$. The solution shows
  behavior in close agreement with typical axonal spiking but is
  derived directly from first principles of electrodynamics and wave
  propagation without any reference to the standard ad-hoc
  reaction-diffusion approach of HH.}
\label{fig:spike}
\end{figure}

This solution to the non-linear problem can be directly
applied to the case of the single axon, using experimentally
measured physiological parameters, thus providing a more precise
characterization of the propagating action potential.  An example
of a numerical solution of equation \cref{eq:ak} for the non-resonant
condition using $\beta_k^\prime = \exp(i\pi/4)$, $\beta_k =
2\exp(-i\pi/4)$, $\alpha_k = 3$, and $\gamma_k = 1.996 + i$ is shown
in \cref{fig:spike}. The solution shows behavior in close agreement
with a typical axonal action potential.

We emphasize that this result was derived from first principles based
on the electrical properties of the axon without the need for an
artificial reaction-diffusion model with multiple adjustable
parameters (thresholds, time constants, etc). In particular, equation
\cref{eq:ak} reveals that the inverse proportionality of frequency and
wave number in the brain wave dispersion relation admits a closed
analytical form of a wave nonlinear equation whose solution is a
persistent traveling axonal nonlinear wave (i.e., the action potential
spike) resulting from the collective non-resonant interactions of
multiple low amplitude wave modes.

We note here that although we have assumed an idealized
perfectly cylindrical model for clarity, the same formalism can be
carried through with more complex geometries.  However, it should
also be recognized that the propagating nonlinear electrodynamic
waves have the capability of deforming the geometry of the charged
membrane, which is consistent with theoretical \cite{El_Hady2015-fz}
and observational \cite{Ling2018-xg} evidence of mechanistic waves
(APPulse \citep{Johnson2018-px,Johnson2018-ye}) that accompany the
action potential propagation.

\subsection{Critical behavior of waves}
\label{sec:crit}

\subsubsection{Critical points}

We have previously demonstrated in 
\citep{Galinsky:2021b,Galinsky:2023a,Galinsky:2023b} that equation
\cref{eq:ak} can be rewritten in terms of a pair of coupled
equations for the amplitude and phase as 
\begin{align}
\label{eq:AC}
\der{A}{t} &=
\gamma A + A^2\left[R_a \cos{(\phi-\Phi)}- \alpha\right],
\\
\label{eq:BC}
\der{\phi}{t} &=\o + A R_\phi \cos{\phi},
\end{align}
where we omitted the subscript $k$ from all variables and assumed
$a(t)=A(t) e^{i\phi(t)}$. The parameters $R_a$, $R_\phi$, and $\Phi$
can be expressed through $\beta$, and $\beta^\prime$ as shown in
\citep{Galinsky:2021b,Galinsky:2023a,Galinsky:2023b}. 

An equilibrium (i.e.,~$dA/dt=d\phi/dt=0$) solution of
\cref{eq:AC,eq:BC} can be found from
\begin{align}
-\frac{\gamma}{\o}R_\phi\cos{\phi} + R_a\cos{(\phi-\Phi)} - \alpha = 0,
\end{align}
with equilibrium values $\phi_e\equiv$ const and
$A_e=-\o/R_\phi\cos{\phi_e}=-\gamma/(R_a\cos{(\phi_e-\Phi)} - \alpha)\equiv$ const.
This shows that for $\alpha > R_a |\sin\Phi|$
there exist critical values $A_c$, $\phi_c$ and $\mu_c$
($\mu=\gamma/\o$) where the equilibrium solution vanishes, such that 
\begin{align}
\mu_c&=
\frac
{R_a\cos{(\phi_c-\Phi)}-\alpha}
{R_\phi\cos{\phi_c}}
\nn
&=
\frac{1}{R_\phi}\left[R_a \cos{\Phi} \pm \sqrt{\alpha^2-(R_a \sin\Phi)^2}\right]
\\
\phi_c&= \arctan\left[\frac{R_a\sin{\Phi}}{R_a\cos{\Phi} -\mu_c R_\phi}\right] 
\nn
&=
\arctan\left[\frac{R_a\sin{\Phi}}{\pm\sqrt{\alpha^2-(R_a\sin{\Phi})^2}}\right],
\\
A_c &=-\frac{\o}{R_\phi\cos{\phi_c}} = - \frac{\gamma}{\mu_c R_\phi\cos{\phi_c}}.
\end{align}

These solutions provide the basis for an analysis of the
critical regimes via a bifurcation analysis.

\subsubsection{Bifurcation analysis}

The standard approach to analyzing the behavior of critical
systems is to linearize the system equations around the critical
point, then determine the stability of the system via the
eigenvalues of the Jacobian (e.g., \cite{Strogatz:2000}). The
linearized system of equations \cref{eq:AC} and \cref{eq:BC}
at the critical point $(A_c,\phi_c)$ results in
\begin{align}
\label{eq:AL}
\der{A}{t} &= \left(\gamma + 2 A_c\left[R_a \cos{(\phi_c-\Phi)}-
  \alpha\right]\right)A \nn 
&-A_c^2 R_a \sin{(\phi_c-\Phi)}\phi,
\\
\label{eq:BL}
\der{\phi}{t} &=R_\phi \cos{(\phi_c)} A - A_c R_\phi \sin{(\phi_c)} \phi.
\end{align}

For different parameter ranges the system \cref{eq:AL,eq:BL} (and
hence the original system \cref{eq:AC,eq:BC} or \cref{eq:ak}) shows
different behavior corresponding to different bifurcation types,
including both the saddle node on an invariant circle (SNIC)
bifurcation (representative for Type I axon spiking) and Hopf
bifurcation (that is claimed to be responsible for Type II axon
spiking) \citep{Prescott2014-mh}.  For
example, taking a limiting case of $R_a\sim \alpha$ with $\Phi=0$
(or $\Phi=\pi$) and $\phi_c=\pi$, the eigenvalues of the Jacobian
matrix become
\begin{align}
\label{eq:L1}
\lambda_1 = 0, \qquad \lambda_2 = \gamma - 2\o\frac{\alpha\pm R_a}{R_\phi},
\end{align}
thus the system undergoes the SNIC bifurcation ($\lambda_1=0$ and
$\lambda_2 < 0$ for for $\mu<2\mu_c$). 

For an alternative limiting case of $R_a\ll \alpha$ with $\Phi=-\pi/2$
and $\phi_c\approx \pi$, the eigenvalues of the Jacobian matrix become 
\begin{align}
\label{eq:L2}
\lambda_{1,2} &= q
\pm
\sqrt{q^2-\o^2 R_a/R_\phi}
\\
q &= \frac{\gamma}{2}-\o\frac{\alpha}{R_\phi}
\end{align}
and in this case for $q=0$
(or $\mu=2\mu_c$) the eigenvalues $\lambda_{1,2}$ are pure imaginary,
crossing the imaginary axis with a change of parameter $\mu$ (either
$\o$ or $\gamma$), which is an example
of a Hopf bifurcation.  Thus, the wave model of action potential
shows that nonlinear axon wave includes multiple critical regimes and
produces different spiking behavior consistent with different
experimentally detected types.

It should be noted that the nonlinear system
{\cref{eq:AC,eq:BC}} is not a simple harmonic oscillator system.
For a harmonic oscillator the amplitude $A$ is constant
(does not change at all) and the phase $\phi$ is changing rapidly with
a constant rate $\o$. The nonlinear system \cref{eq:AC,eq:BC} in the
sub-critical regime, i.e., when $\mu < \mu_c$, shows the oscillations
where the rate of phase change is not constant anymore and the
amplitude $A$ is changing as well, reaching the maximum
$A_{max}=\gamma/(\alpha-R_a)$ and the minimum
$A_{min}=\gamma/(\alpha+R_a)$ for $dA/dt=0$ when $\phi=\Phi$ and
$\phi=\Phi+\pi$ respectively.  

\subsubsection{Spike rate analysis}

As in \citep{Galinsky:2021b,Galinsky:2023a,Galinsky:2023b} we can estimate
the effective period of spiking ${T}_s$ (or its inverse -- either the
firing rate 1/${T}_s$ or the effective firing frequency $2\pi/{T}_s$)
from \cref{eq:BC} by substituting $A$ with either $A_{min}$ (for
positive spikes, $|\phi_c-\Phi|>\pi/2$) or $A_{max}$ (for negative
spikes, $|\phi_c-\Phi|<\pi/2$) as for the most of the time (except for
the short spike duration time) the amplitude $A$ will be close to one
of those values, hence
\begin{align}
{T}_s &= \int\limits_0^{2\pi}\frac{d\phi}{\o + \frac{\gamma R_\phi}{\alpha\pm R_a}\cos{\phi}} 
= \frac{1}{\o}\int\limits_0^{2\pi}\frac{d\phi}{1 + \frac{\mu}{\mu_c}\nu \cos{\phi}} \nn
&= \frac{2\pi}{\o\sqrt{1-\nu^2\mu^2/\mu_c^2}},
\end{align}
where 
\begin{align}
\nu = \frac{R_a\cos{(\phi_c-\Phi)} - \alpha}{(\alpha \pm R_a) \cos\phi_c},
\end{align}
and the
effective firing frequency $\omega_s$
\begin{align}
\omega_s &= \frac{2\pi}{T_s} =\o\sqrt{1-\nu^2\mu^2/\mu_c^2}.
\label{eq:o}
\end{align} 
As in the case discussed above
where $\Phi=0$ (or $\Phi=\pi$) and $\phi_c=\pi$ (also discussed in
\citep{Galinsky:2021b,Galinsky:2023a,Galinsky:2023b}) results in
$\nu=1$, hence gives $\omega_s$=0 when $\mu$ reaches the critical
value $\mu_c$, that is it allows spiking with arbitrary low
frequencies -- the typical behavior of Type I neurons \citep{Prescott2014-mh}.
In the alternative case of $\Phi=-\pi/2$ and
$\phi_c=\pi$, $\nu=\alpha/(\alpha+R_a) < 1$, hence at the critical
point the spiking frequency $\o_s$ can not be less than the minimum
value of $\o\sqrt{1-\nu^2}>0$ -- the behavior attributed to Type II
neurons \citep{Prescott2014-mh}.

\subsubsection{Influence of the applied potential}

Our construction of the conductivity tensor in
\cref{eq:SigmaT} included an adjustable normalization $V$ that
represents the equilibrium voltage drop across the membrane because
the vast majority of experiments investigating neuronal spiking
involve some form of manipulation of $V$, such as ``voltage
clamping''.  From dispersion relation expressions
\cref{eq:dispersion,eq:Sigpp,eq:gamma_d,eq:gamma_e,eq:lam_para} it
follows that
\begin{align}
\mu &= \frac{\gamma_e-\gamma_d}{\o} = \mu_0 +
\frac{\gamma_d}{\omega_0}\left(1-\frac{V_0}{V}\right), 
\label{eq:muV}
\end{align} 
where $\mu_0$ and $\o_0$ are the critical parameter and the linear
wave frequency evaluated at $V=V_0$. Therefore in the sub-critical
($\mu<\mu_c$) regime increasing the voltage difference $V$ across the
membrane, or hyperpolarizing the membrane, increases the criticality
parameter $\mu$, hence decreases the firing frequency $\omega_s$,
stopping the oscillatory (spiking) behavior completely when the
critical point $\mu_c$ is reached. In the super-critical ($\mu>\mu_c$)
case, i.e., when the neuron is not firing, decreasing the voltage
difference $V$ (depolarizing the membrane as it is done in voltage
clamping experiments) decreases the criticality parameter $\mu$ and
makes neuron fire either at non-zero frequency (similar to Type II
neuron) or at arbitrary low frequency (similar to Type I neuron). A
special case of a neuron firing a single spike at the critical point
may also appear if an update of the cross membrane voltage proceeds
too slowly and the system is able to relax back and stay at or above
the critical point, but the periodic firing will emerge with
increasing firing frequency $\omega_s$ when depolarization continues
moving $\mu$ further in the sub-critical range.

\subsubsection{Implications for neural networks}

As shown in
\citep{Galinsky:2021a,Galinsky:2021c,Galinsky:2022a}, the network
organized from such nonlinear oscillators shows synchronization
properties that neither linear oscillators nor diffusive-reactive HH
neurons are capable of producing. Therefore,
the current view that a single neuron can be approximated by the
reactive HH system that is communicating through the cable-like
diffusive signal propagation with other neurons in the networks of
interconnected neurons may not be entirely appropriate for
understanding the dynamics of
brain communication. A more appropriate view may be
to consider that the critical
synchronized state is formed both at a single neuron level and in
their interconnected networks by multiple waves that are constantly
generated at axonal membranes, interact and propagate along those
membrane interfaces, making the networks they form 
to be more appropriately analogous to webs
of highly tensioned strings rather than networks of leaky pipes with
slow diffusive flow of some substance inside those pipes. 

In this ``string theory'' view of neural networks, all the
specific details of the complex biochemical processes that mediate
the membrane voltages are not seen as the actual mechanism behind
axonal spiking nor the subsequent signal propagation in single
neurons and networks of neurons.  Rather, the details about opening
and closing of voltage gated channels, about different number of
Na$^+$, K$^+$, Cl$^-$, Ca$^{2+}$, etc., channels, about differences in
kinetics of those carrier channels, about operation of ATP mediated
carrier pumps, etc., all serve to ``tune'' the membranal
strings by keeping the individual membranes, and hence, the network
as a whole, at or close to the critical level.

\section{Conclusion}

Highly non-linear systems in nature present a significant
problem in data analysis and interpretation because they can produce
a wide variety of seemingly disparate and unrelated coherent
phenomena.  This is particularly true in critical systems where
small parameter variations produced drastically different system
configurations.  Without a physical model for such systems, one is
left with a confusing conglomeration of experimentally observed and
often seemingly contradictory effects without a guiding principle
for understanding the underlying system dynamics.  And without a
guiding theoretical framework, data analysis strategies must often
fall back on essentially \textit{ad hoc} fitting methods.  The more
complex the system, the more parameters are required.  Such
strategies make it possible to fit the data, but deriving a link to
the actual system dynamics in the absense of a theoretical
framework is problematic.

The human brain is a spectacular example of such a non-linear
critical system.  But the lack of a physical theory of brain
activity has led research down that familiar pathway.  So while the
pioneering work of Hodgkin and Huxley \cite{pmid12991237} provided a
new unifying framework for fitting the action potential, it must be
recognized for what it is: an \textit{ad hoc} multiparameteric
fitting method without a physical model.  It is not surprising then
that it has some glaring weaknesses, as noted above, not least of
which is the difficulty in relating the neuronal action potential to
large scale brain network communication.  Nevertheless, it has
remained the standard model for the action potential and forms the
basis for subsequent methods that rely on the empirical fitting of a single measured
axonal signal waveform to a set of \textit{ad hoc} multi-parametric
differential equations with multiple fitting parameters as is
typically employed by a multitude of single-neuron spiking models
\citep{pmid19431309,Nagumo1962,pmid7260316,pmid18244602,Gerstner:2014:NDS:2635959,pmid33192427,pmid33288909}.

Our recent development of a general physical model (WETCOW)
for brain activity derived from the first principles of
electrodynamics
\cite{Galinsky:2020a,Galinsky:2020b,Galinsky:2021a,Galinsky:2021c}
was motivated by the desire to address this problem by constructing
a single unifying framework for understanding brain activity at all
scales, from neuron to network.  Subsequent papers focussed on the
large scale effects such as network synchronization, learning, and
neuronal avalanches
\cite{Galinsky:2022a,Galinsky:2023a,Galinsky:2023b}.  While this
model was developed with all neural tissues in mind, and therefore
implicitly applicable to single neurons, we never explicitly solved
this problem, which simply involved applying the general theory to the
appropriate tissue model of a single neuron.  The objective of this
paper was to solve this problem and therefore explicitly demonstrate
that our general theory works at the range of scales relevant to
brain activity.

In doing so, we have demonstrated this theory of the neuron
action potential is the same that has already demonstrated the ability to explain
multiple observed macroscopic brain electric activity such as
extracellular spiking, efficient brain synchronization, neuronal
avalanches, and memory and learning mechanisms
\cite{Galinsky:2020a,Galinsky:2020b,Galinsky:2021a,Galinsky:2022a,Galinsky:2023a,Galinsky:2023b}
that are not explained by the standard HH model.  We have thus
demonstrated a theory which bridges the gap between the most elemental
brain electrical unit - the neuron, and the large scale collective
synchronous behavior of the brain.

The construction of a physical theory from the first
principles of electrodynamics begged the question of the
relationship to existing electrodynamic models.  The most obvious
candidate is the ubiquitous 'cable theory' which has a long history
in attempted descriptions of neuronal signals.  However, as we
demonstrated in \cref{sec:cable}, it is derivable from our more
general theory but only by the imposition of conditions that limit
its applicability to real neurons.  The cable equations were
subsequently shown to be inadequate to characterize the action
potential under a wide range of realistic conditions.

Recognition that the HH model has never been capable of
solving the problem of characterizing the action potential in
myelinated axons led us to consider that problem within our theory.
We found that the solution was straightforward because our theory
explicitly incorporates both geometrical and physiological tissue
parameters.  This resulted in predictions for wave speeds consistent
with measured values in both myelinated and unmyelinated axons.  It
should be noted that these results have practical significance
because they provide a direct method for relating neuronal activity
to disease states wherein demyelination exists, such as Multiple
Sclerosis \cite{Costa:2023}, and myelin pathogenesis, such as
Alzheimer's Disease \cite{Cai:2016,Maitre:2023}.

The ability of our general WETCOW theory to describe both
spatially extended (including network level) effects as well as
neuron scale effects, led to the demonstration of some remarkable
similarities between the two scales of brain phenomena.  In
\cref{sec:crit} we demonstrated that the critical behaviour
previously shown to be evident in collective synchronous spiking and
neuronal avalanches
\cite{Galinsky:2021b,Galinsky:2023a,Galinsky:2023b}, was similarly
manifest in the neuronal signal where now it corresponds to the
characteristics of Type I and Type II neurons.

One obvious question these results raise is the logic of the current
view of neuronal signaling being \textit{created} by the HH mechanism
of ion exchange, particularly in light of the demonstrated
inadequecy of the diffusion picture. The traveling coherent
non-linear waves predicted by our theory based solely on the
bioelectric properties of the tissues will \textit{cause} a
time-dependent voltage drop across the neuronal membrane that will
influence transmembrane permeability, and therefore have the
ability to open and close multiple voltage gated channels in
synchrony. In this view, the problematic question of how ion
channels mysteriously synchronize to produce an action potential
never arises.

\section*{Acknowledgements}
LRF and VLG were supported by NSF grant ACI-1550405 and NIH grant R01
AG054049. 



\appendix

\section*{Appendix}

\subsection{Normalization constants}
\label{app:norm}
It is convenient to introduce
normalization constants $C_\perp$, $C_\parallel$, and $C$ 
for radial eigenmodes of the linear model as
\begin{align}
\label{eq:phiInt}
C &= \frac{1}{\kappa^2}\!\!\int\limits_{1}^{1+\dl} \!\! r
\left[\der{\phi_{r}}{r}\right]^2 dr = 
\!\!\int\limits_{1}^{1+\dl} \!\! r\phi_{r}^{2} dr \nn
&= \dl\!\!\int\limits_{0}^{1} \!\! (1+\dl r)\phi_{r}^{2} dr \approx
\dl\!\!\int\limits_{0}^{1} \!\! \phi_{r}^{2} dr, \\
C^r_\perp &= \frac{1}{\kappa^2}\!\!\int\limits_{1}^{1+\dl} \!\! r \ln
r \left[\der{\phi_{r}}{r}\right]^2 dr,\\ 
C^z_\perp &= \!\!\int\limits_{1}^{1+\dl} \!\! r \ln r \phi_{r}^{2} dr,\\
C^r_\perp &\approx C^z_\perp 
= 
\dl\!\!\int\limits_{0}^{1} \!\! (1+\dl r)\ln(1+\dl r)
\phi_{r}^{2} dr \approx \frac{\dl}{2}C,\\
C_\parallel &= \!\!\int\limits_{1}^{1+\dl} \!\! \der{(r\ln r)}{r}
\phi_{r}^{2} dr  
\nn
&=
\dl\!\!\int\limits_{0}^{1} \!\! (1+\ln(1+\dl r)) \phi_{r}^{2}
dr \approx C.
\end{align}

\subsection{Normalized coefficients}
\label{app:coef}

It is also convenient to introduce normalization coefficient
$D_\perp$, $D_\parallel$, and $D$ for the nonlinear model as
\begin{align}
D_\parallel &= -\frac{\Sigma_{zz}^{\prime}}{2
  C\varkappa^2}\!\!\int\limits_{1}^{1+\dl} \!\!
r \phi_{r}^3 dr,\nn 
\label{eq:D}
D_\perp &=\frac{
  \Sigma_{rr}^{\prime}}{C \varkappa^2}\!\!\int\limits_{1}^{1+\dl}
\!\! r\phi_{r}\left[\der{\phi_{r}}{r}\right]^2 dr,\\ 
D &=\frac{(2 \Sigma_{zr}^\prime-\Sigma_{rz}^{\prime})}{6
  C \varkappa^2}\!\!\int\limits_{1}^{1+\dl} \!\!
\phi_{r}^3 dr, 
\nonumber
\end{align}
where $\varkappa^2$ is given by {\cref{eq:varkappa}}.  and the
normalization parameters $C_\perp$, $C_\parallel$, and $C$ are
provided in \Cref{app:norm} by \cref{eq:phiInt}.

\begin{thebibliography}{50}%
\makeatletter
\providecommand \@ifxundefined [1]{%
 \@ifx{#1\undefined}
}%
\providecommand \@ifnum [1]{%
 \ifnum #1\expandafter \@firstoftwo
 \else \expandafter \@secondoftwo
 \fi
}%
\providecommand \@ifx [1]{%
 \ifx #1\expandafter \@firstoftwo
 \else \expandafter \@secondoftwo
 \fi
}%
\providecommand \natexlab [1]{#1}%
\providecommand \enquote  [1]{``#1''}%
\providecommand \bibnamefont  [1]{#1}%
\providecommand \bibfnamefont [1]{#1}%
\providecommand \citenamefont [1]{#1}%
\providecommand \href@noop [0]{\@secondoftwo}%
\providecommand \href [0]{\begingroup \@sanitize@url \@href}%
\providecommand \@href[1]{\@@startlink{#1}\@@href}%
\providecommand \@@href[1]{\endgroup#1\@@endlink}%
\providecommand \@sanitize@url [0]{\catcode `\\12\catcode `\$12\catcode
  `\&12\catcode `\#12\catcode `\^12\catcode `\_12\catcode `\%12\relax}%
\providecommand \@@startlink[1]{}%
\providecommand \@@endlink[0]{}%
\providecommand \url  [0]{\begingroup\@sanitize@url \@url }%
\providecommand \@url [1]{\endgroup\@href {#1}{\urlprefix }}%
\providecommand \urlprefix  [0]{URL }%
\providecommand \Eprint [0]{\href }%
\providecommand \doibase [0]{https://doi.org/}%
\providecommand \selectlanguage [0]{\@gobble}%
\providecommand \bibinfo  [0]{\@secondoftwo}%
\providecommand \bibfield  [0]{\@secondoftwo}%
\providecommand \translation [1]{[#1]}%
\providecommand \BibitemOpen [0]{}%
\providecommand \bibitemStop [0]{}%
\providecommand \bibitemNoStop [0]{.\EOS\space}%
\providecommand \EOS [0]{\spacefactor3000\relax}%
\providecommand \BibitemShut  [1]{\csname bibitem#1\endcsname}%
\let\auto@bib@innerbib\@empty
\bibitem [{\citenamefont {Hodgkin}\ and\ \citenamefont
  {Huxley}(1952)}]{pmid12991237}%
  \BibitemOpen
  \bibfield  {author} {\bibinfo {author} {\bibfnamefont {A.~L.}\ \bibnamefont
  {Hodgkin}}\ and\ \bibinfo {author} {\bibfnamefont {A.~F.}\ \bibnamefont
  {Huxley}},\ }\bibfield  {title} {\bibinfo {title} {{{A} quantitative
  description of membrane current and its application to conduction and
  excitation in nerve}},\ }\href@noop {} {\bibfield  {journal} {\bibinfo
  {journal} {J. Physiol. (Lond.)}\ }\textbf {\bibinfo {volume} {117}},\
  \bibinfo {pages} {500} (\bibinfo {year} {1952})}\BibitemShut {NoStop}%
\bibitem [{\citenamefont {Fitzhugh}(1961)}]{pmid19431309}%
  \BibitemOpen
  \bibfield  {author} {\bibinfo {author} {\bibfnamefont {R.}~\bibnamefont
  {Fitzhugh}},\ }\bibfield  {title} {\bibinfo {title} {{{I}mpulses and
  {P}hysiological {S}tates in {T}heoretical {M}odels of {N}erve {M}embrane}},\
  }\href@noop {} {\bibfield  {journal} {\bibinfo  {journal} {Biophys. J.}\
  }\textbf {\bibinfo {volume} {1}},\ \bibinfo {pages} {445} (\bibinfo {year}
  {1961})}\BibitemShut {NoStop}%
\bibitem [{\citenamefont {Nagumo}\ \emph {et~al.}(1962)\citenamefont {Nagumo},
  \citenamefont {Arimoto},\ and\ \citenamefont {Yoshizawa}}]{Nagumo1962}%
  \BibitemOpen
  \bibfield  {author} {\bibinfo {author} {\bibfnamefont {J.}~\bibnamefont
  {Nagumo}}, \bibinfo {author} {\bibfnamefont {S.}~\bibnamefont {Arimoto}},\
  and\ \bibinfo {author} {\bibfnamefont {S.}~\bibnamefont {Yoshizawa}},\
  }\bibfield  {title} {\bibinfo {title} {An active pulse transmission line
  simulating nerve axon},\ }\href {https://doi.org/10.1109/jrproc.1962.288235}
  {\bibfield  {journal} {\bibinfo  {journal} {Proceedings of the {IRE}}\
  }\textbf {\bibinfo {volume} {50}},\ \bibinfo {pages} {2061} (\bibinfo {year}
  {1962})}\BibitemShut {NoStop}%
\bibitem [{\citenamefont {Morris}\ and\ \citenamefont
  {Lecar}(1981)}]{pmid7260316}%
  \BibitemOpen
  \bibfield  {author} {\bibinfo {author} {\bibfnamefont {C.}~\bibnamefont
  {Morris}}\ and\ \bibinfo {author} {\bibfnamefont {H.}~\bibnamefont {Lecar}},\
  }\bibfield  {title} {\bibinfo {title} {{{V}oltage oscillations in the
  barnacle giant muscle fiber}},\ }\href@noop {} {\bibfield  {journal}
  {\bibinfo  {journal} {Biophys. J.}\ }\textbf {\bibinfo {volume} {35}},\
  \bibinfo {pages} {193} (\bibinfo {year} {1981})}\BibitemShut {NoStop}%
\bibitem [{\citenamefont {Izhikevich}(2003)}]{pmid18244602}%
  \BibitemOpen
  \bibfield  {author} {\bibinfo {author} {\bibfnamefont {E.~M.}\ \bibnamefont
  {Izhikevich}},\ }\bibfield  {title} {\bibinfo {title} {{{S}imple model of
  spiking neurons}},\ }\href@noop {} {\bibfield  {journal} {\bibinfo  {journal}
  {IEEE Trans Neural Netw}\ }\textbf {\bibinfo {volume} {14}},\ \bibinfo
  {pages} {1569} (\bibinfo {year} {2003})}\BibitemShut {NoStop}%
\bibitem [{\citenamefont {Buzsaki}(2006)}]{buzsaki2006rhythms}%
  \BibitemOpen
  \bibfield  {author} {\bibinfo {author} {\bibfnamefont {G.}~\bibnamefont
  {Buzsaki}},\ }\href
  {https://urldefense.com/v3/__https://books.google.com/books?id=ldz58irprjYC__;!!Mih3wA!VDUnBPduahevOhOiOqYHTZZHh4H7dA_b9_L4xp0hCJOVds515-icRlSbEfQA$}
  {\emph {\bibinfo {title} {Rhythms of the Brain}}}\ (\bibinfo  {publisher}
  {Oxford University Press},\ \bibinfo {year} {2006})\BibitemShut {NoStop}%
\bibitem [{\citenamefont {Strassberg}\ and\ \citenamefont
  {DeFelice}(1993)}]{Strassberg1993-nh}%
  \BibitemOpen
  \bibfield  {author} {\bibinfo {author} {\bibfnamefont {A.~F.}\ \bibnamefont
  {Strassberg}}\ and\ \bibinfo {author} {\bibfnamefont {L.~J.}\ \bibnamefont
  {DeFelice}},\ }\bibfield  {title} {\bibinfo {title} {{Limitations of the
  Hodgkin-Huxley formalism: Effects of single channel kinetics on transmembrane
  voltage dynamics}},\ }\href@noop {} {\bibfield  {journal} {\bibinfo
  {journal} {Neural Comput.}\ }\textbf {\bibinfo {volume} {5}},\ \bibinfo
  {pages} {843} (\bibinfo {year} {1993})}\BibitemShut {NoStop}%
\bibitem [{\citenamefont {Meunier}\ and\ \citenamefont
  {Segev}(2002)}]{Meunier2002-rs}%
  \BibitemOpen
  \bibfield  {author} {\bibinfo {author} {\bibfnamefont {C.}~\bibnamefont
  {Meunier}}\ and\ \bibinfo {author} {\bibfnamefont {I.}~\bibnamefont
  {Segev}},\ }\bibfield  {title} {\bibinfo {title} {Playing the devil's
  advocate: is the {Hodgkin-Huxley} model useful?},\ }\href@noop {} {\bibfield
  {journal} {\bibinfo  {journal} {Trends Neurosci.}\ }\textbf {\bibinfo
  {volume} {25}},\ \bibinfo {pages} {558} (\bibinfo {year} {2002})}\BibitemShut
  {NoStop}%
\bibitem [{\citenamefont {Yamazaki}\ \emph {et~al.}(2022)\citenamefont
  {Yamazaki}, \citenamefont {Vo-Ho}, \citenamefont {Bulsara},\ and\
  \citenamefont {Le}}]{Yamazaki2022-cm}%
  \BibitemOpen
  \bibfield  {author} {\bibinfo {author} {\bibfnamefont {K.}~\bibnamefont
  {Yamazaki}}, \bibinfo {author} {\bibfnamefont {V.-K.}\ \bibnamefont {Vo-Ho}},
  \bibinfo {author} {\bibfnamefont {D.}~\bibnamefont {Bulsara}},\ and\ \bibinfo
  {author} {\bibfnamefont {N.}~\bibnamefont {Le}},\ }\bibfield  {title}
  {\bibinfo {title} {Spiking neural networks and their applications: A
  review},\ }\href@noop {} {\bibfield  {journal} {\bibinfo  {journal} {Brain
  Sci.}\ }\textbf {\bibinfo {volume} {12}},\ \bibinfo {pages} {863} (\bibinfo
  {year} {2022})}\BibitemShut {NoStop}%
\bibitem [{\citenamefont {Gerstner}\ \emph {et~al.}(2014)\citenamefont
  {Gerstner}, \citenamefont {Kistler}, \citenamefont {Naud},\ and\
  \citenamefont {Paninski}}]{Gerstner:2014:NDS:2635959}%
  \BibitemOpen
  \bibfield  {author} {\bibinfo {author} {\bibfnamefont {W.}~\bibnamefont
  {Gerstner}}, \bibinfo {author} {\bibfnamefont {W.~M.}\ \bibnamefont
  {Kistler}}, \bibinfo {author} {\bibfnamefont {R.}~\bibnamefont {Naud}},\ and\
  \bibinfo {author} {\bibfnamefont {L.}~\bibnamefont {Paninski}},\ }\href@noop
  {} {\emph {\bibinfo {title} {Neuronal Dynamics: From Single Neurons to
  Networks and Models of Cognition}}}\ (\bibinfo  {publisher} {Cambridge
  University Press},\ \bibinfo {address} {New York, NY, USA},\ \bibinfo {year}
  {2014})\BibitemShut {NoStop}%
\bibitem [{\citenamefont {Kulkarni}\ \emph {et~al.}(2020)\citenamefont
  {Kulkarni}, \citenamefont {Ranft},\ and\ \citenamefont
  {Hakim}}]{pmid33192427}%
  \BibitemOpen
  \bibfield  {author} {\bibinfo {author} {\bibfnamefont {A.}~\bibnamefont
  {Kulkarni}}, \bibinfo {author} {\bibfnamefont {J.}~\bibnamefont {Ranft}},\
  and\ \bibinfo {author} {\bibfnamefont {V.}~\bibnamefont {Hakim}},\ }\bibfield
   {title} {\bibinfo {title} {{{S}ynchronization, {S}tochasticity, and {P}hase
  {W}aves in {N}euronal {N}etworks {W}ith {S}patially-{S}tructured
  {C}onnectivity}},\ }\href@noop {} {\bibfield  {journal} {\bibinfo  {journal}
  {Front Comput Neurosci}\ }\textbf {\bibinfo {volume} {14}},\ \bibinfo {pages}
  {569644} (\bibinfo {year} {2020})}\BibitemShut {NoStop}%
\bibitem [{\citenamefont {Kim}\ and\ \citenamefont
  {Sejnowski}(2021)}]{pmid33288909}%
  \BibitemOpen
  \bibfield  {author} {\bibinfo {author} {\bibfnamefont {R.}~\bibnamefont
  {Kim}}\ and\ \bibinfo {author} {\bibfnamefont {T.~J.}\ \bibnamefont
  {Sejnowski}},\ }\bibfield  {title} {\bibinfo {title} {{{S}trong inhibitory
  signaling underlies stable temporal dynamics and working memory in spiking
  neural networks}},\ }\href@noop {} {\bibfield  {journal} {\bibinfo  {journal}
  {Nat Neurosci}\ }\textbf {\bibinfo {volume} {24}},\ \bibinfo {pages} {129}
  (\bibinfo {year} {2021})}\BibitemShut {NoStop}%
\bibitem [{\citenamefont {Galinsky}\ and\ \citenamefont
  {Frank}(2023{\natexlab{a}})}]{Galinsky:2022a}%
  \BibitemOpen
  \bibfield  {author} {\bibinfo {author} {\bibfnamefont {V.~L.}\ \bibnamefont
  {Galinsky}}\ and\ \bibinfo {author} {\bibfnamefont {L.~R.}\ \bibnamefont
  {Frank}},\ }\bibfield  {title} {\bibinfo {title} {{Critically synchronized
  brain waves form an effective, robust and flexible basis for human memory and
  learning}},\ }\href {https://doi.org/10.1038/s41598-023-31365-6} {\bibfield
  {journal} {\bibinfo  {journal} {Sci Rep}\ }\textbf {\bibinfo {volume} {13}},\
  \bibinfo {pages} {4343} (\bibinfo {year} {2023}{\natexlab{a}})}\BibitemShut
  {NoStop}%
\bibitem [{\citenamefont {Galinsky}\ and\ \citenamefont
  {Frank}(2020{\natexlab{a}})}]{Galinsky:2020a}%
  \BibitemOpen
  \bibfield  {author} {\bibinfo {author} {\bibfnamefont {V.~L.}\ \bibnamefont
  {Galinsky}}\ and\ \bibinfo {author} {\bibfnamefont {L.~R.}\ \bibnamefont
  {Frank}},\ }\bibfield  {title} {\bibinfo {title} {Universal theory of brain
  waves: from linear loops to nonlinear synchronized spiking and collective
  brain rhythms},\ }\href@noop {} {\bibfield  {journal} {\bibinfo  {journal}
  {Physical Review Research}\ }\textbf {\bibinfo {volume} {2}},\ \bibinfo
  {pages} {023061} (\bibinfo {year} {2020}{\natexlab{a}})}\BibitemShut
  {NoStop}%
\bibitem [{\citenamefont {Galinsky}\ and\ \citenamefont
  {Frank}(2020{\natexlab{b}})}]{Galinsky:2020b}%
  \BibitemOpen
  \bibfield  {author} {\bibinfo {author} {\bibfnamefont {V.~L.}\ \bibnamefont
  {Galinsky}}\ and\ \bibinfo {author} {\bibfnamefont {L.~R.}\ \bibnamefont
  {Frank}},\ }\bibfield  {title} {\bibinfo {title} {{Brain Waves: Emergence of
  Localized, Persistent, Weakly Evanescent Cortical Loops}},\ }\href@noop {}
  {\bibfield  {journal} {\bibinfo  {journal} {J. of Cognitive Neurosci}\
  }\textbf {\bibinfo {volume} {32}},\ \bibinfo {pages} {2178} (\bibinfo {year}
  {2020}{\natexlab{b}})}\BibitemShut {NoStop}%
\bibitem [{\citenamefont {Galinsky}\ and\ \citenamefont
  {Frank}(2021{\natexlab{a}})}]{Galinsky:2021a}%
  \BibitemOpen
  \bibfield  {author} {\bibinfo {author} {\bibfnamefont {V.~L.}\ \bibnamefont
  {Galinsky}}\ and\ \bibinfo {author} {\bibfnamefont {L.~R.}\ \bibnamefont
  {Frank}},\ }\bibfield  {title} {\bibinfo {title} {Collective synchronous
  spiking in a brain network of coupled nonlinear oscillators},\ }\href
  {https://doi.org/10.1103/PhysRevLett.126.158102} {\bibfield  {journal}
  {\bibinfo  {journal} {Phys. Rev. Lett.}\ }\textbf {\bibinfo {volume} {126}},\
  \bibinfo {pages} {158102} (\bibinfo {year} {2021}{\natexlab{a}})}\BibitemShut
  {NoStop}%
\bibitem [{\citenamefont {Galinsky}\ and\ \citenamefont
  {Frank}(2021{\natexlab{b}})}]{Galinsky:2021c}%
  \BibitemOpen
  \bibfield  {author} {\bibinfo {author} {\bibfnamefont {V.~L.}\ \bibnamefont
  {Galinsky}}\ and\ \bibinfo {author} {\bibfnamefont {L.~R.}\ \bibnamefont
  {Frank}},\ }\bibfield  {title} {\bibinfo {title} {Collective synchronous
  spiking in a brain network of coupled nonlinear oscillators},\ }\href@noop {}
  {\bibfield  {journal} {\bibinfo  {journal} {eprint arXiv:2104.02171}\ }
  (\bibinfo {year} {2021}{\natexlab{b}})}\BibitemShut {NoStop}%
\bibitem [{\citenamefont {Galinsky}\ and\ \citenamefont
  {Frank}(2021{\natexlab{c}})}]{Galinsky:2021b}%
  \BibitemOpen
  \bibfield  {author} {\bibinfo {author} {\bibfnamefont {V.~L.}\ \bibnamefont
  {Galinsky}}\ and\ \bibinfo {author} {\bibfnamefont {L.~R.}\ \bibnamefont
  {Frank}},\ }\bibfield  {title} {\bibinfo {title} {Neuronal avalanches and
  critical dynamics of brain waves},\ }\bibfield  {journal} {\bibinfo
  {journal} {eprint arXiv:2111.07479}\ }\href@noop {} {} (\bibinfo {year}
  {2021}{\natexlab{c}})\BibitemShut {NoStop}%
\bibitem [{\citenamefont {Galinsky}\ and\ \citenamefont
  {Frank}(2023{\natexlab{b}})}]{Galinsky:2023a}%
  \BibitemOpen
  \bibfield  {author} {\bibinfo {author} {\bibfnamefont {V.~L.}\ \bibnamefont
  {Galinsky}}\ and\ \bibinfo {author} {\bibfnamefont {L.~R.}\ \bibnamefont
  {Frank}},\ }\bibfield  {title} {\bibinfo {title} {Critical brain wave
  dynamics of neuronal avalanches},\ }\href@noop {} {\bibfield  {journal}
  {\bibinfo  {journal} {{Frontiers in Physics}}\ }\textbf {\bibinfo {volume}
  {11}} (\bibinfo {year} {2023}{\natexlab{b}})}\BibitemShut {NoStop}%
\bibitem [{\citenamefont {Galinsky}\ and\ \citenamefont
  {Frank}(2023{\natexlab{c}})}]{Galinsky:2023b}%
  \BibitemOpen
  \bibfield  {author} {\bibinfo {author} {\bibfnamefont {V.~L.}\ \bibnamefont
  {Galinsky}}\ and\ \bibinfo {author} {\bibfnamefont {L.~R.}\ \bibnamefont
  {Frank}},\ }\bibfield  {title} {\bibinfo {title} {Neuronal avalanches:
  sandpiles of self organized criticality or critical dynamics of brain
  waves?},\ }\href {https://doi.org/10.1007/s11467-023-1273-7} {\bibfield
  {journal} {\bibinfo  {journal} {{Frontiers of Physics}}\ }\textbf {\bibinfo
  {volume} {18}},\ \bibinfo {pages} {45301} (\bibinfo {year}
  {2023}{\natexlab{c}})}\BibitemShut {NoStop}%
\bibitem [{\citenamefont {Scott}(1975)}]{Scott1975-ht}%
  \BibitemOpen
  \bibfield  {author} {\bibinfo {author} {\bibfnamefont {A.~C.}\ \bibnamefont
  {Scott}},\ }\bibfield  {title} {\bibinfo {title} {The electrophysics of a
  nerve fiber},\ }\href@noop {} {\bibfield  {journal} {\bibinfo  {journal}
  {Rev. Mod. Phys.}\ }\textbf {\bibinfo {volume} {47}},\ \bibinfo {pages} {487}
  (\bibinfo {year} {1975})}\BibitemShut {NoStop}%
\bibitem [{\citenamefont {B{\'e}dard}\ \emph {et~al.}(2004)\citenamefont
  {B{\'e}dard}, \citenamefont {Kr{\"o}ger},\ and\ \citenamefont
  {Destexhe}}]{Bedard2004-lz}%
  \BibitemOpen
  \bibfield  {author} {\bibinfo {author} {\bibfnamefont {C.}~\bibnamefont
  {B{\'e}dard}}, \bibinfo {author} {\bibfnamefont {H.}~\bibnamefont
  {Kr{\"o}ger}},\ and\ \bibinfo {author} {\bibfnamefont {A.}~\bibnamefont
  {Destexhe}},\ }\bibfield  {title} {\bibinfo {title} {Modeling extracellular
  field potentials and the frequency-filtering properties of extracellular
  space},\ }\href@noop {} {\bibfield  {journal} {\bibinfo  {journal} {Biophys.
  J.}\ }\textbf {\bibinfo {volume} {86}},\ \bibinfo {pages} {1829} (\bibinfo
  {year} {2004})}\BibitemShut {NoStop}%
\bibitem [{\citenamefont {Gillespie}\ and\ \citenamefont
  {Stein}(1983)}]{Gillespie1983-qd}%
  \BibitemOpen
  \bibfield  {author} {\bibinfo {author} {\bibfnamefont {M.~J.}\ \bibnamefont
  {Gillespie}}\ and\ \bibinfo {author} {\bibfnamefont {R.~B.}\ \bibnamefont
  {Stein}},\ }\bibfield  {title} {\bibinfo {title} {The relationship between
  axon diameter, myelin thickness and conduction velocity during atrophy of
  mammalian peripheral nerves},\ }\href@noop {} {\bibfield  {journal} {\bibinfo
   {journal} {Brain Res.}\ }\textbf {\bibinfo {volume} {259}},\ \bibinfo
  {pages} {41} (\bibinfo {year} {1983})}\BibitemShut {NoStop}%
\bibitem [{\citenamefont {Arancibia-C{\'a}rcamo}\ \emph
  {et~al.}(2017)\citenamefont {Arancibia-C{\'a}rcamo}, \citenamefont {Ford},
  \citenamefont {Cossell}, \citenamefont {Ishida}, \citenamefont {Tohyama},\
  and\ \citenamefont {Attwell}}]{Arancibia-Carcamo2017-pe}%
  \BibitemOpen
  \bibfield  {author} {\bibinfo {author} {\bibfnamefont {I.~L.}\ \bibnamefont
  {Arancibia-C{\'a}rcamo}}, \bibinfo {author} {\bibfnamefont {M.~C.}\
  \bibnamefont {Ford}}, \bibinfo {author} {\bibfnamefont {L.}~\bibnamefont
  {Cossell}}, \bibinfo {author} {\bibfnamefont {K.}~\bibnamefont {Ishida}},
  \bibinfo {author} {\bibfnamefont {K.}~\bibnamefont {Tohyama}},\ and\ \bibinfo
  {author} {\bibfnamefont {D.}~\bibnamefont {Attwell}},\ }\bibfield  {title}
  {\bibinfo {title} {Node of ranvier length as a potential regulator of
  myelinated axon conduction speed},\ }\href@noop {} {\bibfield  {journal}
  {\bibinfo  {journal} {Elife}\ }\textbf {\bibinfo {volume} {6}} (\bibinfo
  {year} {2017})}\BibitemShut {NoStop}%
\bibitem [{\citenamefont {Waxman}\ and\ \citenamefont
  {Melker}(1971)}]{Waxman1971-fm}%
  \BibitemOpen
  \bibfield  {author} {\bibinfo {author} {\bibfnamefont {S.~G.}\ \bibnamefont
  {Waxman}}\ and\ \bibinfo {author} {\bibfnamefont {R.~J.}\ \bibnamefont
  {Melker}},\ }\bibfield  {title} {\bibinfo {title} {Closely spaced nodes of
  ranvier in the mammalian brain},\ }\href@noop {} {\bibfield  {journal}
  {\bibinfo  {journal} {Brain Res.}\ }\textbf {\bibinfo {volume} {32}},\
  \bibinfo {pages} {445} (\bibinfo {year} {1971})}\BibitemShut {NoStop}%
\bibitem [{\citenamefont {Siegel}\ and\ \citenamefont
  {Sapru}(2005)}]{Siegel2005-dh}%
  \BibitemOpen
  \bibfield  {author} {\bibinfo {author} {\bibfnamefont {A.}~\bibnamefont
  {Siegel}}\ and\ \bibinfo {author} {\bibfnamefont {H.~N.}\ \bibnamefont
  {Sapru}},\ }\bibinfo {title} {Essential neuroscience}\ (\bibinfo  {publisher}
  {Lippincott Williams and Wilkins},\ \bibinfo {address} {Philadelphia, PA},\
  \bibinfo {year} {2005})\ p.\ \bibinfo {pages} {257}\BibitemShut {NoStop}%
\bibitem [{\citenamefont {Hodgkin}\ \emph {et~al.}(1946)\citenamefont
  {Hodgkin}, \citenamefont {Rushton},\ and\ \citenamefont
  {Adrian}}]{Hodgkin:1946}%
  \BibitemOpen
  \bibfield  {author} {\bibinfo {author} {\bibfnamefont {A.~L.}\ \bibnamefont
  {Hodgkin}}, \bibinfo {author} {\bibfnamefont {W.~A.~H.}\ \bibnamefont
  {Rushton}},\ and\ \bibinfo {author} {\bibfnamefont {E.~D.}\ \bibnamefont
  {Adrian}},\ }\bibfield  {title} {\bibinfo {title} {The electrical constants
  of a crustacean nerve fibre},\ }\href
  {https://doi.org/10.1098/rspb.1946.0024} {\bibfield  {journal} {\bibinfo
  {journal} {Proceedings of the Royal Society of London. Series B - Biological
  Sciences}\ }\textbf {\bibinfo {volume} {133}},\ \bibinfo {pages} {444}
  (\bibinfo {year} {1946})}\BibitemShut {NoStop}%
\bibitem [{\citenamefont {Rall}(1962)}]{Rall:1962}%
  \BibitemOpen
  \bibfield  {author} {\bibinfo {author} {\bibfnamefont {W.}~\bibnamefont
  {Rall}},\ }\bibfield  {title} {\bibinfo {title} {Theory of physiological
  properties of dendrites},\ }\href@noop {} {\bibfield  {journal} {\bibinfo
  {journal} {Annals of the New York Academy of Sciences}\ }\textbf {\bibinfo
  {volume} {96}},\ \bibinfo {pages} {1071} (\bibinfo {year}
  {1962})}\BibitemShut {NoStop}%
\bibitem [{\citenamefont {Rall}(2011)}]{Rall2011-hr}%
  \BibitemOpen
  \bibfield  {author} {\bibinfo {author} {\bibfnamefont {W.}~\bibnamefont
  {Rall}},\ }\bibinfo {title} {Core conductor theory and cable properties of
  neurons},\ in\ \href@noop {} {\emph {\bibinfo {booktitle} {Comprehensive
  Physiology}}}\ (\bibinfo  {publisher} {John Wiley \& Sons, Inc.},\ \bibinfo
  {address} {Hoboken, NJ, USA},\ \bibinfo {year} {2011})\BibitemShut {NoStop}%
\bibitem [{\citenamefont {Segev}\ and\ \citenamefont
  {London}(2000)}]{Segev:2000}%
  \BibitemOpen
  \bibfield  {author} {\bibinfo {author} {\bibfnamefont {I.}~\bibnamefont
  {Segev}}\ and\ \bibinfo {author} {\bibfnamefont {M.}~\bibnamefont {London}},\
  }\bibfield  {title} {\bibinfo {title} {Untangling dendrites with quantitative
  models},\ }\href@noop {} {\bibfield  {journal} {\bibinfo  {journal}
  {Science}\ }\textbf {\bibinfo {volume} {290}},\ \bibinfo {pages} {744}
  (\bibinfo {year} {2000})}\BibitemShut {NoStop}%
\bibitem [{\citenamefont {Holmes}(2014)}]{Holmes2014-pl}%
  \BibitemOpen
  \bibfield  {author} {\bibinfo {author} {\bibfnamefont {W.~R.}\ \bibnamefont
  {Holmes}},\ }\bibfield  {title} {\bibinfo {title} {Cable equation},\ }in\
  \href@noop {} {\emph {\bibinfo {booktitle} {Encyclopedia of Computational
  Neuroscience}}}\ (\bibinfo  {publisher} {Springer New York},\ \bibinfo
  {address} {New York, NY},\ \bibinfo {year} {2014})\ pp.\ \bibinfo {pages}
  {1--13}\BibitemShut {NoStop}%
\bibitem [{\citenamefont {Pagkalos}\ \emph {et~al.}(2023)\citenamefont
  {Pagkalos}, \citenamefont {Chavlis},\ and\ \citenamefont
  {Poirazi}}]{Pagkalos2023}%
  \BibitemOpen
  \bibfield  {author} {\bibinfo {author} {\bibfnamefont {M.}~\bibnamefont
  {Pagkalos}}, \bibinfo {author} {\bibfnamefont {S.}~\bibnamefont {Chavlis}},\
  and\ \bibinfo {author} {\bibfnamefont {P.}~\bibnamefont {Poirazi}},\
  }\bibfield  {title} {\bibinfo {title} {Introducing the dendrify framework for
  incorporating dendrites to spiking neural networks},\ }\href
  {https://doi.org/10.1038/s41467-022-35747-8} {\bibfield  {journal} {\bibinfo
  {journal} {Nature Communications}\ }\textbf {\bibinfo {volume} {14}},\
  \bibinfo {pages} {131} (\bibinfo {year} {2023})}\BibitemShut {NoStop}%
\bibitem [{\citenamefont {Spruston}\ \emph {et~al.}(2013)\citenamefont
  {Spruston}, \citenamefont {Häusser},\ and\ \citenamefont
  {Stuart}}]{Spruston:2013}%
  \BibitemOpen
  \bibfield  {author} {\bibinfo {author} {\bibfnamefont {N.}~\bibnamefont
  {Spruston}}, \bibinfo {author} {\bibfnamefont {M.}~\bibnamefont {Häusser}},\
  and\ \bibinfo {author} {\bibfnamefont {G.}~\bibnamefont {Stuart}},\
  }\bibfield  {title} {\bibinfo {title} {Chapter 11 - information processing in
  dendrites and spines},\ }in\ \href
  {https://doi.org/https://doi.org/10.1016/B978-0-12-385870-2.00011-1} {\emph
  {\bibinfo {booktitle} {Fundamental Neuroscience (Fourth Edition)}}},\
  \bibinfo {editor} {edited by\ \bibinfo {editor} {\bibfnamefont {L.~R.}\
  \bibnamefont {Squire}}, \bibinfo {editor} {\bibfnamefont {D.}~\bibnamefont
  {Berg}}, \bibinfo {editor} {\bibfnamefont {F.~E.}\ \bibnamefont {Bloom}},
  \bibinfo {editor} {\bibfnamefont {S.}~\bibnamefont {{du Lac}}}, \bibinfo
  {editor} {\bibfnamefont {A.}~\bibnamefont {Ghosh}},\ and\ \bibinfo {editor}
  {\bibfnamefont {N.~C.}\ \bibnamefont {Spitzer}}}\ (\bibinfo  {publisher}
  {Academic Press},\ \bibinfo {address} {San Diego},\ \bibinfo {year} {2013})\
  \bibinfo {edition} {fourth edition}\ ed.,\ pp.\ \bibinfo {pages}
  {231--260}\BibitemShut {NoStop}%
\bibitem [{\citenamefont {Cohen}\ \emph {et~al.}(2020)\citenamefont {Cohen},
  \citenamefont {Popovic}, \citenamefont {Klooster}, \citenamefont {Weil},
  \citenamefont {M{\"o}bius}, \citenamefont {Nave},\ and\ \citenamefont
  {Kole}}]{Cohen2020-ie}%
  \BibitemOpen
  \bibfield  {author} {\bibinfo {author} {\bibfnamefont {C.~C.~H.}\
  \bibnamefont {Cohen}}, \bibinfo {author} {\bibfnamefont {M.~A.}\ \bibnamefont
  {Popovic}}, \bibinfo {author} {\bibfnamefont {J.}~\bibnamefont {Klooster}},
  \bibinfo {author} {\bibfnamefont {M.-T.}\ \bibnamefont {Weil}}, \bibinfo
  {author} {\bibfnamefont {W.}~\bibnamefont {M{\"o}bius}}, \bibinfo {author}
  {\bibfnamefont {K.-A.}\ \bibnamefont {Nave}},\ and\ \bibinfo {author}
  {\bibfnamefont {M.~H.~P.}\ \bibnamefont {Kole}},\ }\bibfield  {title}
  {\bibinfo {title} {Saltatory conduction along myelinated axons involves a
  periaxonal nanocircuit},\ }\href@noop {} {\bibfield  {journal} {\bibinfo
  {journal} {Cell}\ }\textbf {\bibinfo {volume} {180}},\ \bibinfo {pages} {311}
  (\bibinfo {year} {2020})}\BibitemShut {NoStop}%
\bibitem [{\citenamefont {Lim}\ and\ \citenamefont
  {Rasband}(2020)}]{Lim2020-sq}%
  \BibitemOpen
  \bibfield  {author} {\bibinfo {author} {\bibfnamefont {B.~C.}\ \bibnamefont
  {Lim}}\ and\ \bibinfo {author} {\bibfnamefont {M.~N.}\ \bibnamefont
  {Rasband}},\ }\bibfield  {title} {\bibinfo {title} {Saltatory conduction:
  Jumping to new conclusions},\ }\href@noop {} {\bibfield  {journal} {\bibinfo
  {journal} {Curr. Biol.}\ }\textbf {\bibinfo {volume} {30}},\ \bibinfo {pages}
  {R326} (\bibinfo {year} {2020})}\BibitemShut {NoStop}%
\bibitem [{\citenamefont {Sharmin}\ \emph {et~al.}(2023)\citenamefont
  {Sharmin}, \citenamefont {Karal}, \citenamefont {Mahbub},\ and\ \citenamefont
  {Rabbani}}]{Sharmin2023-sf}%
  \BibitemOpen
  \bibfield  {author} {\bibinfo {author} {\bibfnamefont {S.}~\bibnamefont
  {Sharmin}}, \bibinfo {author} {\bibfnamefont {M.~A.~S.}\ \bibnamefont
  {Karal}}, \bibinfo {author} {\bibfnamefont {Z.~B.}\ \bibnamefont {Mahbub}},\
  and\ \bibinfo {author} {\bibfnamefont {K.~S.-E.}\ \bibnamefont {Rabbani}},\
  }\bibfield  {title} {\bibinfo {title} {Increase in conduction velocity in
  myelinated nerves due to stretch - an experimental verification},\
  }\href@noop {} {\bibfield  {journal} {\bibinfo  {journal} {Front. Neurosci.}\
  }\textbf {\bibinfo {volume} {17}},\ \bibinfo {pages} {1084004} (\bibinfo
  {year} {2023})}\BibitemShut {NoStop}%
\bibitem [{\citenamefont {Sengupta}\ \emph {et~al.}(2013)\citenamefont
  {Sengupta}, \citenamefont {Faisal}, \citenamefont {Laughlin},\ and\
  \citenamefont {Niven}}]{Sengupta2013-ya}%
  \BibitemOpen
  \bibfield  {author} {\bibinfo {author} {\bibfnamefont {B.}~\bibnamefont
  {Sengupta}}, \bibinfo {author} {\bibfnamefont {A.~A.}\ \bibnamefont
  {Faisal}}, \bibinfo {author} {\bibfnamefont {S.~B.}\ \bibnamefont
  {Laughlin}},\ and\ \bibinfo {author} {\bibfnamefont {J.~E.}\ \bibnamefont
  {Niven}},\ }\bibfield  {title} {\bibinfo {title} {The effect of cell size and
  channel density on neuronal information encoding and energy efficiency},\
  }\href@noop {} {\bibfield  {journal} {\bibinfo  {journal} {J. Cereb. Blood
  Flow Metab.}\ }\textbf {\bibinfo {volume} {33}},\ \bibinfo {pages} {1465}
  (\bibinfo {year} {2013})}\BibitemShut {NoStop}%
\bibitem [{\citenamefont {Hu}\ and\ \citenamefont {Jonas}(2014)}]{Hu2014-lj}%
  \BibitemOpen
  \bibfield  {author} {\bibinfo {author} {\bibfnamefont {H.}~\bibnamefont
  {Hu}}\ and\ \bibinfo {author} {\bibfnamefont {P.}~\bibnamefont {Jonas}},\
  }\bibfield  {title} {\bibinfo {title} {A supercritical density of na(+)
  channels ensures fast signaling in {GABAergic} interneuron axons},\
  }\href@noop {} {\bibfield  {journal} {\bibinfo  {journal} {Nat. Neurosci.}\
  }\textbf {\bibinfo {volume} {17}},\ \bibinfo {pages} {686} (\bibinfo {year}
  {2014})}\BibitemShut {NoStop}%
\bibitem [{\citenamefont {Freeman}\ \emph {et~al.}(2016)\citenamefont
  {Freeman}, \citenamefont {Desmazi{\`e}res}, \citenamefont {Fricker},
  \citenamefont {Lubetzki},\ and\ \citenamefont {Sol-Foulon}}]{Freeman2016-ph}%
  \BibitemOpen
  \bibfield  {author} {\bibinfo {author} {\bibfnamefont {S.~A.}\ \bibnamefont
  {Freeman}}, \bibinfo {author} {\bibfnamefont {A.}~\bibnamefont
  {Desmazi{\`e}res}}, \bibinfo {author} {\bibfnamefont {D.}~\bibnamefont
  {Fricker}}, \bibinfo {author} {\bibfnamefont {C.}~\bibnamefont {Lubetzki}},\
  and\ \bibinfo {author} {\bibfnamefont {N.}~\bibnamefont {Sol-Foulon}},\
  }\bibfield  {title} {\bibinfo {title} {Mechanisms of sodium channel
  clustering and its influence on axonal impulse conduction},\ }\href@noop {}
  {\bibfield  {journal} {\bibinfo  {journal} {Cell. Mol. Life Sci.}\ }\textbf
  {\bibinfo {volume} {73}},\ \bibinfo {pages} {723} (\bibinfo {year}
  {2016})}\BibitemShut {NoStop}%
\bibitem [{\citenamefont {Arhem}\ \emph {et~al.}(2006)\citenamefont {Arhem},
  \citenamefont {Klement},\ and\ \citenamefont {Blomberg}}]{Arhem2006-dl}%
  \BibitemOpen
  \bibfield  {author} {\bibinfo {author} {\bibfnamefont {P.}~\bibnamefont
  {Arhem}}, \bibinfo {author} {\bibfnamefont {G.}~\bibnamefont {Klement}},\
  and\ \bibinfo {author} {\bibfnamefont {C.}~\bibnamefont {Blomberg}},\
  }\bibfield  {title} {\bibinfo {title} {Channel density regulation of firing
  patterns in a cortical neuron model},\ }\href@noop {} {\bibfield  {journal}
  {\bibinfo  {journal} {Biophys. J.}\ }\textbf {\bibinfo {volume} {90}},\
  \bibinfo {pages} {4392} (\bibinfo {year} {2006})}\BibitemShut {NoStop}%
\bibitem [{\citenamefont {El~Hady}\ and\ \citenamefont
  {Machta}(2015)}]{El_Hady2015-fz}%
  \BibitemOpen
  \bibfield  {author} {\bibinfo {author} {\bibfnamefont {A.}~\bibnamefont
  {El~Hady}}\ and\ \bibinfo {author} {\bibfnamefont {B.~B.}\ \bibnamefont
  {Machta}},\ }\bibfield  {title} {\bibinfo {title} {Mechanical surface waves
  accompany action potential propagation},\ }\href@noop {} {\bibfield
  {journal} {\bibinfo  {journal} {Nat. Commun.}\ }\textbf {\bibinfo {volume}
  {6}},\ \bibinfo {pages} {6697} (\bibinfo {year} {2015})}\BibitemShut
  {NoStop}%
\bibitem [{\citenamefont {Ling}\ \emph {et~al.}(2018)\citenamefont {Ling},
  \citenamefont {Boyle}, \citenamefont {Goetz}, \citenamefont {Zhou},
  \citenamefont {Quan}, \citenamefont {Alfonso}, \citenamefont {Huang},\ and\
  \citenamefont {Palanker}}]{Ling2018-xg}%
  \BibitemOpen
  \bibfield  {author} {\bibinfo {author} {\bibfnamefont {T.}~\bibnamefont
  {Ling}}, \bibinfo {author} {\bibfnamefont {K.~C.}\ \bibnamefont {Boyle}},
  \bibinfo {author} {\bibfnamefont {G.}~\bibnamefont {Goetz}}, \bibinfo
  {author} {\bibfnamefont {P.}~\bibnamefont {Zhou}}, \bibinfo {author}
  {\bibfnamefont {Y.}~\bibnamefont {Quan}}, \bibinfo {author} {\bibfnamefont
  {F.~S.}\ \bibnamefont {Alfonso}}, \bibinfo {author} {\bibfnamefont {T.~W.}\
  \bibnamefont {Huang}},\ and\ \bibinfo {author} {\bibfnamefont
  {D.}~\bibnamefont {Palanker}},\ }\bibfield  {title} {\bibinfo {title}
  {Full-field interferometric imaging of propagating action potentials},\
  }\href@noop {} {\bibfield  {journal} {\bibinfo  {journal} {Light Sci. Appl.}\
  }\textbf {\bibinfo {volume} {7}},\ \bibinfo {pages} {107} (\bibinfo {year}
  {2018})}\BibitemShut {NoStop}%
\bibitem [{\citenamefont {Johnson}\ and\ \citenamefont
  {Winlow}(2018{\natexlab{a}})}]{Johnson2018-px}%
  \BibitemOpen
  \bibfield  {author} {\bibinfo {author} {\bibfnamefont {A.~S.}\ \bibnamefont
  {Johnson}}\ and\ \bibinfo {author} {\bibfnamefont {W.}~\bibnamefont
  {Winlow}},\ }\bibfield  {title} {\bibinfo {title} {The soliton and the action
  potential -- primary elements underlying sentience},\ }\href@noop {}
  {\bibfield  {journal} {\bibinfo  {journal} {Front. Physiol.}\ }\textbf
  {\bibinfo {volume} {9}} (\bibinfo {year} {2018}{\natexlab{a}})}\BibitemShut
  {NoStop}%
\bibitem [{\citenamefont {Johnson}\ and\ \citenamefont
  {Winlow}(2018{\natexlab{b}})}]{Johnson2018-ye}%
  \BibitemOpen
  \bibfield  {author} {\bibinfo {author} {\bibfnamefont {A.}~\bibnamefont
  {Johnson}}\ and\ \bibinfo {author} {\bibfnamefont {W.}~\bibnamefont
  {Winlow}},\ }\bibfield  {title} {\bibinfo {title} {Mysteries of the action
  potential},\ }\href@noop {} {\bibfield  {journal} {\bibinfo  {journal}
  {Physiology News}\ ,\ \bibinfo {pages} {38}} (\bibinfo {year}
  {2018}{\natexlab{b}})}\BibitemShut {NoStop}%
\bibitem [{\citenamefont {Strogatz}(2000)}]{Strogatz:2000}%
  \BibitemOpen
  \bibfield  {author} {\bibinfo {author} {\bibfnamefont {S.~H.}\ \bibnamefont
  {Strogatz}},\ }\href@noop {} {\emph {\bibinfo {title} {{Nonlinear Dynamics
  and Chaos: With Applications to Physics, Biology, Chemistry and
  Engineering}}}}\ (\bibinfo  {publisher} {Westview Press},\ \bibinfo {year}
  {2000})\BibitemShut {NoStop}%
\bibitem [{\citenamefont {Prescott}(2014)}]{Prescott2014-mh}%
  \BibitemOpen
  \bibfield  {author} {\bibinfo {author} {\bibfnamefont {S.~A.}\ \bibnamefont
  {Prescott}},\ }\bibfield  {title} {\bibinfo {title} {Excitability: Types i,
  {II}, and {III}},\ }in\ \href@noop {} {\emph {\bibinfo {booktitle}
  {Encyclopedia of Computational Neuroscience}}}\ (\bibinfo  {publisher}
  {Springer New York},\ \bibinfo {address} {New York, NY},\ \bibinfo {year}
  {2014})\ pp.\ \bibinfo {pages} {1--7}\BibitemShut {NoStop}%
\bibitem [{\citenamefont {Coutinho~Costa}\ \emph {et~al.}(2023)\citenamefont
  {Coutinho~Costa}, \citenamefont {Araújo}, \citenamefont {Alves-Leon},\ and\
  \citenamefont {Gomes}}]{Costa:2023}%
  \BibitemOpen
  \bibfield  {author} {\bibinfo {author} {\bibfnamefont {V.~G.}\ \bibnamefont
  {Coutinho~Costa}}, \bibinfo {author} {\bibfnamefont {S.~E.-S.}\ \bibnamefont
  {Araújo}}, \bibinfo {author} {\bibfnamefont {S.~V.}\ \bibnamefont
  {Alves-Leon}},\ and\ \bibinfo {author} {\bibfnamefont {F.~C.~A.}\
  \bibnamefont {Gomes}},\ }\bibfield  {title} {\bibinfo {title} {Central
  nervous system demyelinating diseases: glial cells at the hub of pathology},\
  }\href@noop {} {\bibfield  {journal} {\bibinfo  {journal} {Frontiers in
  Immunology}\ }\textbf {\bibinfo {volume} {14}} (\bibinfo {year}
  {2023})}\BibitemShut {NoStop}%
\bibitem [{\citenamefont {Cai}\ and\ \citenamefont {Xiao}(2016)}]{Cai:2016}%
  \BibitemOpen
  \bibfield  {author} {\bibinfo {author} {\bibfnamefont {Z.}~\bibnamefont
  {Cai}}\ and\ \bibinfo {author} {\bibfnamefont {M.}~\bibnamefont {Xiao}},\
  }\bibfield  {title} {\bibinfo {title} {Oligodendrocytes and alzheimer's
  disease},\ }\href@noop {} {\bibfield  {journal} {\bibinfo  {journal}
  {International Journal of Neuroscience}\ }\textbf {\bibinfo {volume} {126}},\
  \bibinfo {pages} {97} (\bibinfo {year} {2016})}\BibitemShut {NoStop}%
\bibitem [{\citenamefont {Maitre}\ \emph {et~al.}(2023)\citenamefont {Maitre},
  \citenamefont {Jeltsch-David}, \citenamefont {Okechukwu}, \citenamefont
  {Klein}, \citenamefont {Patte-Mensah},\ and\ \citenamefont
  {Mensah-Nyagan}}]{Maitre:2023}%
  \BibitemOpen
  \bibfield  {author} {\bibinfo {author} {\bibfnamefont {M.}~\bibnamefont
  {Maitre}}, \bibinfo {author} {\bibfnamefont {H.}~\bibnamefont
  {Jeltsch-David}}, \bibinfo {author} {\bibfnamefont {N.~G.}\ \bibnamefont
  {Okechukwu}}, \bibinfo {author} {\bibfnamefont {C.}~\bibnamefont {Klein}},
  \bibinfo {author} {\bibfnamefont {C.}~\bibnamefont {Patte-Mensah}},\ and\
  \bibinfo {author} {\bibfnamefont {A.-G.}\ \bibnamefont {Mensah-Nyagan}},\
  }\bibfield  {title} {\bibinfo {title} {Myelin in alzheimer's disease: culprit
  or bystander?},\ }\href@noop {} {\bibfield  {journal} {\bibinfo  {journal}
  {Acta Neuropathologica Communications}\ }\textbf {\bibinfo {volume} {11}},\
  \bibinfo {pages} {56} (\bibinfo {year} {2023})}\BibitemShut {NoStop}%
\bibitem [{\citenamefont {Fraser}\ \emph {et~al.}(2019)\citenamefont {Fraser},
  \citenamefont {Horgan}, \citenamefont {Miller}, \citenamefont {Johnson},\
  and\ \citenamefont {Winlow}}]{Fraser2019-gi}%
  \BibitemOpen
  \bibfield  {author} {\bibinfo {author} {\bibfnamefont {J.}~\bibnamefont
  {Fraser}}, \bibinfo {author} {\bibfnamefont {R.}~\bibnamefont {Horgan}},
  \bibinfo {author} {\bibfnamefont {D.}~\bibnamefont {Miller}}, \bibinfo
  {author} {\bibfnamefont {A.}~\bibnamefont {Johnson}},\ and\ \bibinfo {author}
  {\bibfnamefont {B.}~\bibnamefont {Winlow}},\ }\bibfield  {title} {\bibinfo
  {title} {On the topic of mysteries of the action potential},\ }\href@noop {}
  {\bibfield  {journal} {\bibinfo  {journal} {Physiology News}\ ,\ \bibinfo
  {pages} {6}} (\bibinfo {year} {2019})}\BibitemShut {NoStop}%
\end{thebibliography}

%
\end{document}